# User-Centered Insights into Assistive Navigation Technologies for Individuals with Visual Impairment

**Iman Soltani[1], Johnaton Schofield[1], Mehran Madani[2], Daniel Kish[3] and Parisa Emami-Naeini[4]**

### Abstract

Navigational challenges significantly impact the independence and mobility of **I**ndividuals with **V**isual **I**mpairment (IVI). While numerous assistive technologies exist, their adoption remains limited due to usability challenges, financial constraints, and a lack of alignment with user needs. This study employs a mixed-methods approach, combining structured surveys and virtual workshops with 19 IVI to investigate their experiences, needs, and preferences regarding assistive technologies for navigation and daily living. The survey results provide insights into participants' technological competence, preferences for assistive devices, and willingness to adopt new solutions. In parallel, workshop discussions offer qualitative perspectives on key navigation challenges, including difficulties in detecting overhead obstacles, navigating environments with complex layout, and the limitations of existing technologies. Findings highlight the need for assistive devices that integrate both navigational guidance and high-level spatial awareness, allowing users to build mental maps of their surroundings. Additionally, multimodal feedback, combining audio, haptic, and tactile cues, emerges as a crucial feature to accommodate diverse user preferences and environmental conditions. The study also underscores financial and training barriers that limit access to advanced assistive technologies. Based on these insights, we recommend the development of customizable, user-friendly, and most importantly affordable navigation aids that align with the daily needs of IVI. The findings from this study provide guidance for technology developers, researchers, and policymakers working toward more inclusive and effective assistive solutions.

## Introduction

Effective navigation is crucial for the independence and quality of life of individuals with visual impairment (IVI). The ability to move safely and confidently through various environments enhances their day-to-day activities and supports social integration and autonomy. However, navigating spaces designed primarily for the sighted can pose significant challenges, often leading to social isolation and reduced participation in community life Rokach et al. (2021). As such, understanding and improving the navigational aids available to IVI is essential for fostering equal opportunities and supporting their full participation in society. This study examines the adoption and effectiveness of technological solutions designed to assist with navigation, directly from the perspective of IVI. It further explores their needs and preferences to inform the development of more user-centered assistive technologies that enhance mobility and independence.

Individuals with Visual Impairment encounter a range of navigation challenges. In outdoor environments, they must navigate through obstacles such as poorly maintained sidewalks, absent audible signals at crossings, overhead obstacles like tree branches, and non-standardized public transportation systems. Indoor navigation presents its own set of hurdles, often exacerbated by the variety in architectural design, the varying and unpredictable arrangement of obstacles, the arbitrary positioning of service centers like front desks, and non-standard placements of entrances, exits, seating areas, trash bins, escalators, stairways, restrooms and elevators. Both settings typically lack the necessary non-visual cues that may facilitate safe and easy navigation, further compounded by the dynamic movement of sighted individuals, who often lack awareness of the challenges faced by those with visual impairments. The general lack of public awareness can further lead to inadequate support and accommodation, especially in public settings. Such barriers not only restrict mobility but also affect confidence, leading to decreased social engagement and increased reliance on assistance. Consequently, these challenges can limit access to employment, education, and essential services Dias et al. (2015); Bakali El Mohamadi et al. (2024); Okolo et al. (2024); for the Blind (2024).

[1]Department of Mechanical and Aerospace Engineering, University of California- Davis, Davis, CA, USA
[2]Department of Housing and Community Development, Sacramento, CA, USA
[3]World Access for the Blind, Placentia, CA, USA
[4]Department of Opthalmology and Visual Sciences, Tschannen Eye Institute, University of California-Davis, Sacramento, CA, USA

**Corresponding author:**
Iman Soltani, Department of Mechanical and Aerospace Engineering, One Shields Ave, Davis, CA, USA.
Email: isoltani@ucdavis.edu

A variety of navigation technologies have been developed to assist the IVI. These technologies range from traditional tools like the white cane equipped with electronic sensors Okolo et al. (2024), to GPS-based apps tailored for the IVI El-taher et al. (2021), wearable devices that use haptic feedback Bineeth Kuriakose and Sandnes (2022), and AI-driven systems that interpret live camera feeds to describe surroundings audibly Fernandes et al. (2019). Research has also explored the potential of virtual reality to train IVI in spatial navigation by simulating real-world environments in a controlled setting Ricci et al. (2024).

Despite advances in navigational assistive technologies, there remains a notable gap in widespread adoption and consistent use of these innovations. Previous studies have acknowledged factors such as high costs Isazade (2023), complexity of use Okolo et al. (2024), and inadequate infrastructure El-taher et al. (2021). However, many of these technology-driven studies fail to closely incorporate the firsthand perspectives and input of IVI. This lack of meaningful engagement with those who have lived experience results in only a partial understanding of the barriers they face, often resulting in technologically advanced solutions that are misaligned with the IVI's day-to-day experiences and challenges. Research indicates that a lack of user-centered design and insufficient involvement of IVI in the development process lead to low adoption rates Ortiz-Escobar et al. (2023). Therefore, a comprehensive understanding that comes from direct user involvement is still lacking.

Our study addresses this gap by examining the experiences, preferences, and needs of individuals with visual impairments (IVI) in the context of navigation. Using a mixed-methods approach, combining virtual workshops for open-ended qualitative discussions with structured surveys for quantitative insights, we aim to build a deeper understanding of the challenges IVI face and the factors that shape their interactions with both traditional and emerging navigation technologies. These insights can inform future efforts to develop solutions that better support independence, mobility, and quality of life for the IVI community.

## Methods

### *Survey*

To ensure that participants' responses were not influenced by the discussions in the workshops, the survey was administered prior to these sessions. This approach also encouraged participants to provide additional thought to the issues they may have not considered before, thus allowing for more focused and productive discussions during the workshop sessions. The survey comprised 27 questions, excluding contact and demographical information, designed to capture a broad spectrum of information, including personal experiences with navigation challenges, usage patterns of assistive technologies, and preferences for feedback types. The questions were a mix of open-ended, and Likert scale, allowing for a deeper understanding of participants' needs and experiences. For instance, participants were asked about their familiarity with Braille, the types of smart devices they use, and their comfort with using various technological aids. They also provided insights into the specific navigation aids they have used, their satisfaction with these technologies, and the financial costs associated with their use. The complete list of the survey questions are available in section .

The survey data analysis employed a dual approach to handle both the open-ended and multiple-choice responses. For qualitative data from open-ended questions, we utilized the GPT-4o language model OpenAI (2024) to perform thematic analysis, which allowed us to extract common themes and significant points efficiently and objectively. This AI-driven approach facilitated the distillation of patterns and common narratives from the qualitative responses. To ensure the accuracy and reliability of this AI-assisted analysis, the authors carefully reviewed the identified themes against the original responses. For quantitative data from the Likert scale questions, as well as some of the thematic analysis outcomes, various visual representations, like graphs and charts, were adopted to clearly demonstrate the statistics and potential relationships between different variables. This helped in understanding the broader trends and preferences among participants. Additionally, given the extensive range of questions, responses to certain related questions were combined to provide a concise and coherent presentation.

This study protocol was approved by the Institutional Review Board at the University of California-Davis. The study was conducted in accordance with the tenets of the Declaration of Helsinki and the Health Insurance Portability and Accountability Act.

### *Virtual workshops*

Following the survey, the study organized a series of five virtual workshops conducted between June 2023 and April 2024 to engage directly with IVI and gather their insights on navigation technologies, their experiences, and challenges they are facing. These sessions were hosted on the Zoom platform, with a total of 19 individuals attending. Each workshop was facilitated by one of the investigators who acted as the moderator, guiding the discussion through a set of predefined themes while also allowing for open-ended discussions to capture novel ideas and personal experiences. All co-investigators, representing a multidisciplinary team with expertise in design, human-machine interaction, ophthalmology, and the training of IVI for functional independence in daily living, actively participated in these sessions, fostering a comprehensive exchange of ideas and insights. Each of these workshops lasted between one to two hours, depending on the flow of discussion, and covered topics.

Participants were recruited through the UC Davis Department of Ophthalmology and the Sacramento Society for the Blind, which assisted in distributing a digital flier outlining the workshop's intentions and research objectives via email to all its members. Interested volunteers responded directly to our research coordinator. We included adult participants over the age of 18 with extremely low-vision. Of the 19 participants, 17 reported vision equal to or less than light perception in both eyes. Level of visual impairment was self-reported by the participants and was not independently verified.

Each virtual workshop followed a structured, yet flexible procedure. The sessions began with an introduction of the research team and mission, followed by an invitation for participants to introduce themselves. Discussions then moved through a series of predetermined topics: 1) exploring the main navigation challenges faced by participants, 2) the type of information they found most useful for navigation, 3) their preferences for receiving feedback from assistive devices (audio, tactile, haptic, or a combination), and 4) their overall perceptions of existing assistive technologies. The discussions were not rigidly confined to these topics; the moderator allowed the conversation to deviate naturally to encourage brainstorming and uncover novel concepts or issues. All sessions were recorded, and the audio files were subsequently transcribed using the OpenAI's Whisper open-source tool Radford et al. (2023). The transcriptions were analyzed using the GPT-4o language model OpenAI (2024) to extract common and important themes and pointers, ensuring that key insights were accurately and objectively identified and documented for further analysis. The authors carefully validated the extracted content by reviewing the transcripts and cross-referencing them with the outputs of the language model, ensuring that no inaccuracies occurred and that all extracted information and summaries were precise and reliable.

# Results

## Survey Results and Analysis

In the following, we present the analysis of the survey responses, highlighting key themes and insights derived from participant feedback.

### *Braille Literacy and Technological Competence*

This section examines participants' familiarity with Braille and smart devices, providing context for interpreting their navigation-related responses. The questions are as follows:

Q1- Are you familiar with Braille? Yes, no, somewhat.

Q2- What types of **smart devices** do you regularly use? (Smartphone, tablet, personal computer, smartwatches, etc.).

Q3- How technologically competent would you describe yourself? (1-5, from (1) not competent at all using modern smart devices to (5) extremely competent using modern smart devices).

Q4- How likely are you to try new technologies or smart devices? (1-5, from (1) hesitant to try new technologies and prefer to use my existing devices to (5) extremely eager to try new technologies)

Q5- How comfortable are you with using technology and assistive devices to improve independence? (1-5, from (1) not confident at all to (5) very confident)

**Responses to Q1** Among the 19 participants, 10 reported being familiar with Braille, 6 were somewhat familiar, and only 3 had no prior experience.

**Responses to Q2 and Q3**

Figure 1, provides a stacked bar chart, combining participants' self-reported technological competence levels and the types of devices they use, illustrating the distribution of device usage across different competence levels. All 19 participants reported using smartphones, and 14 reported using personal computers or laptops. For these two categories, competence levels ranged from 2 to 5, although their majority identified as technologically competent at level 3 or higher. Only three participants reported using smart visually impaired specialty devices, all of whom were in competence levels 3 and 4.

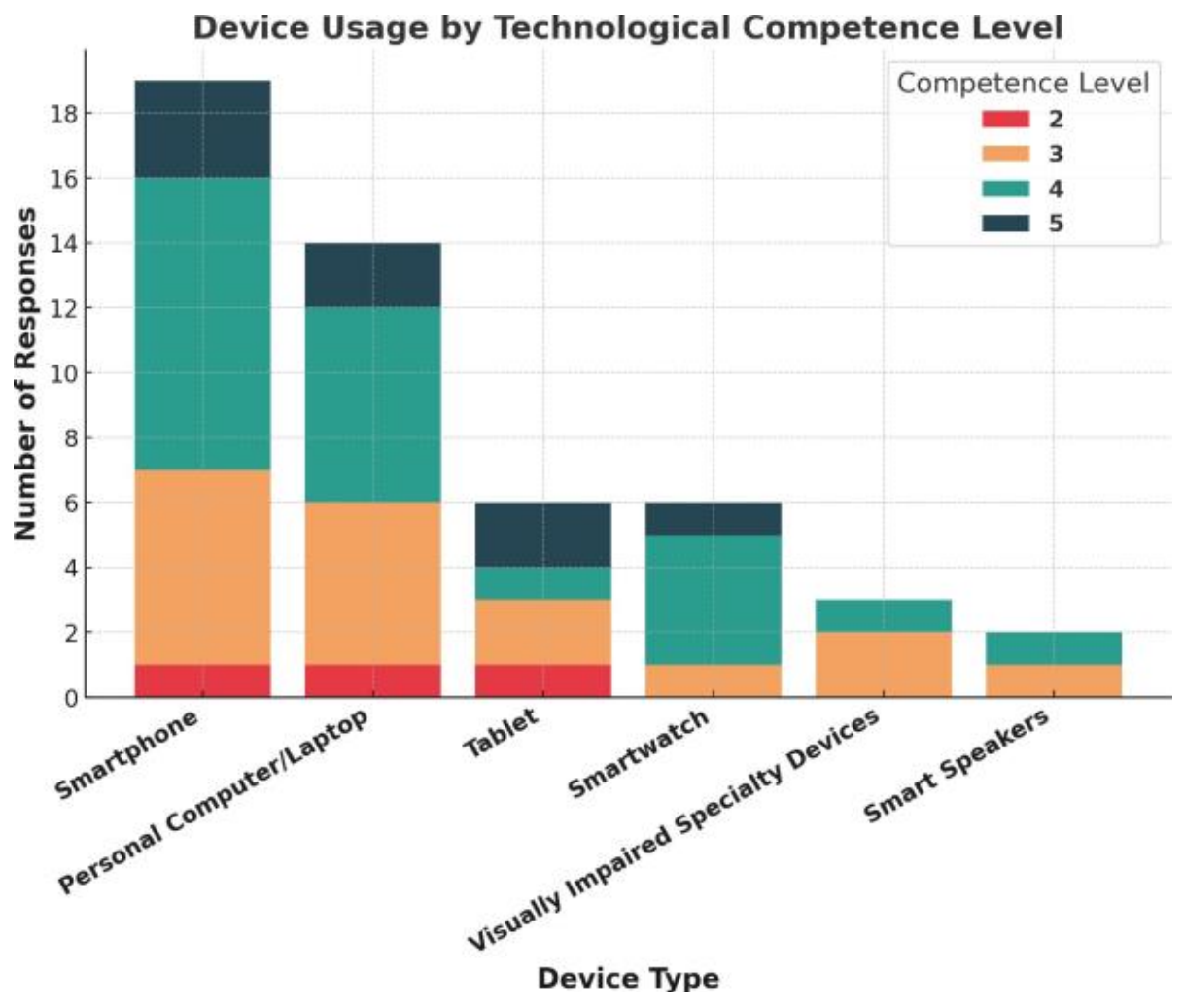

**Figure 1.** Summary of the responses to survey questions Q2 and Q3

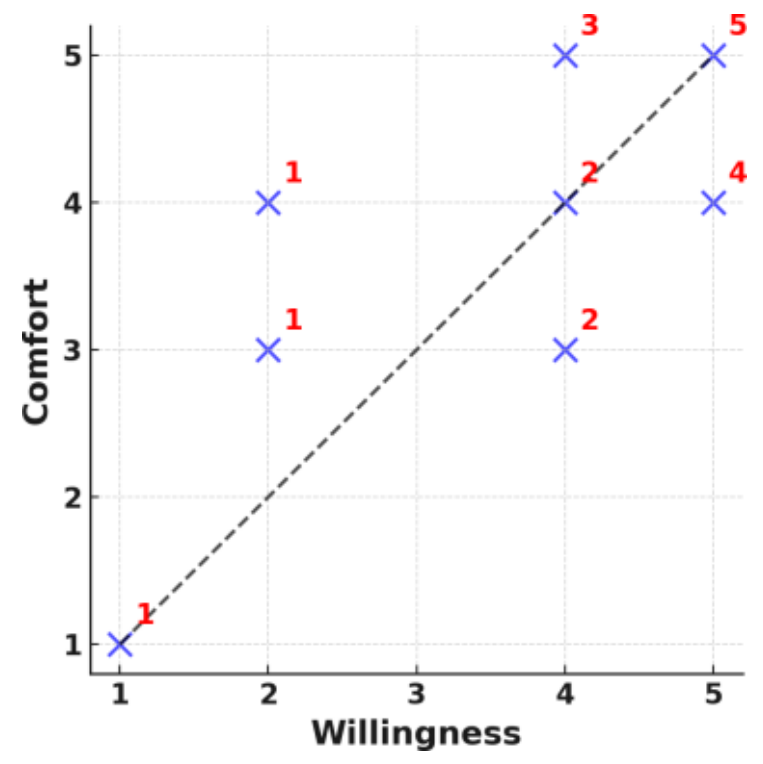

**Figure 2.** Summary of the responses to survey questions Q4 and Q5. The number of participants corresponding to each data point is indicated in red next to each entry.

**Responses to Q4 and Q5**

Figure 2 presents a scatter plot illustrating the relationship between participants' willingness to try new technologies and their comfort with using technology to enhance independence. The number of participants corresponding to each data point is indicated next to each entry.

Among the 19 participants, 18 reported feeling comfortable using technology and assistive devices to improve their independence, with most selecting a rating of 3 or higher. Similarly, many participants (16) expressed a high willingness to try new technologies, rating their eagerness at 4 or above. Not surprisingly, the level of comfort correlates well with the willingness to try new technologies, with a correlation coefficient of 0.737.

### *Assistive Technology Usage*

Q6- Because of your visual impairment, do you use a technology to assist you in performing your daily activities?

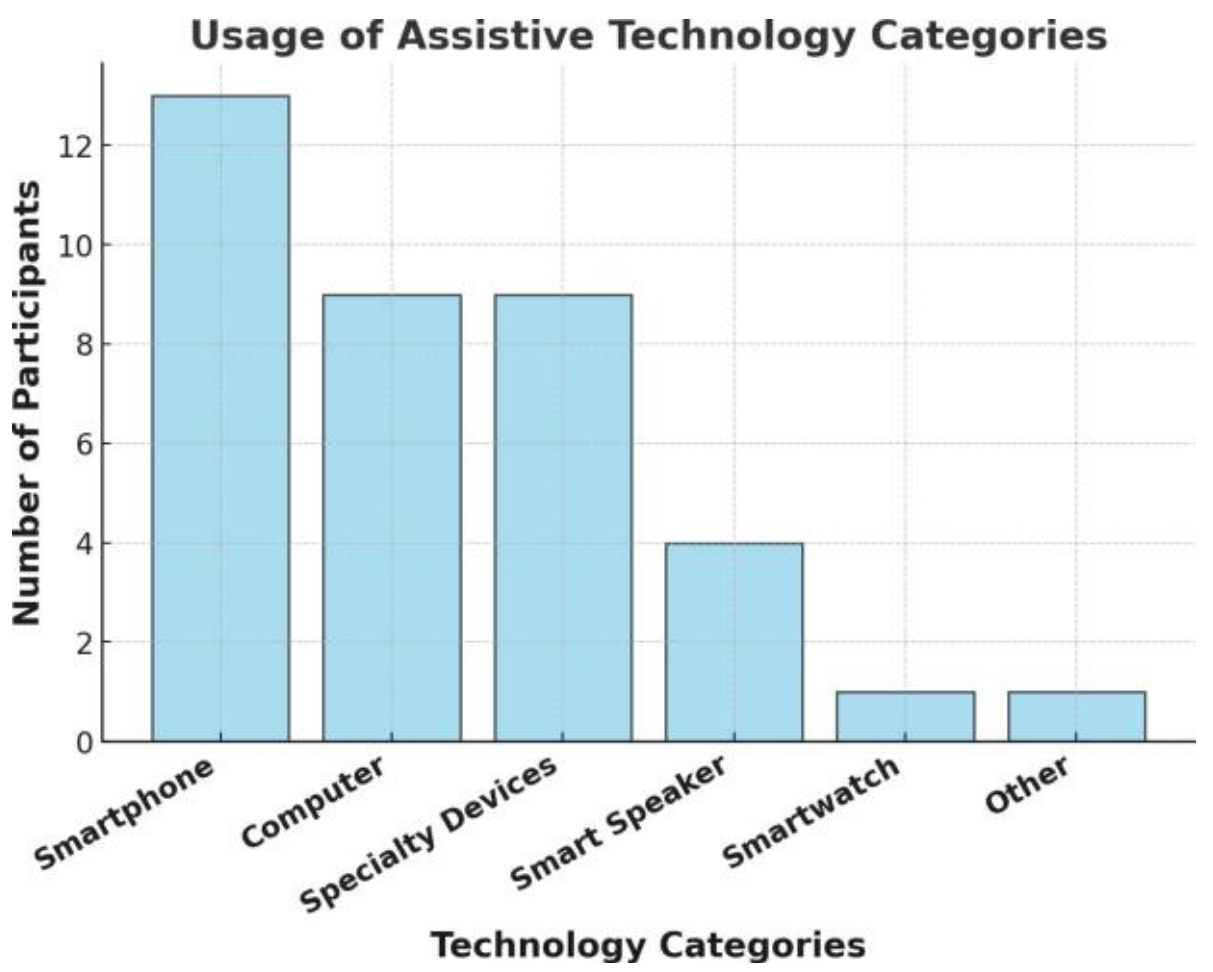

**Figure 3.** Summary of the responses to survey question Q6

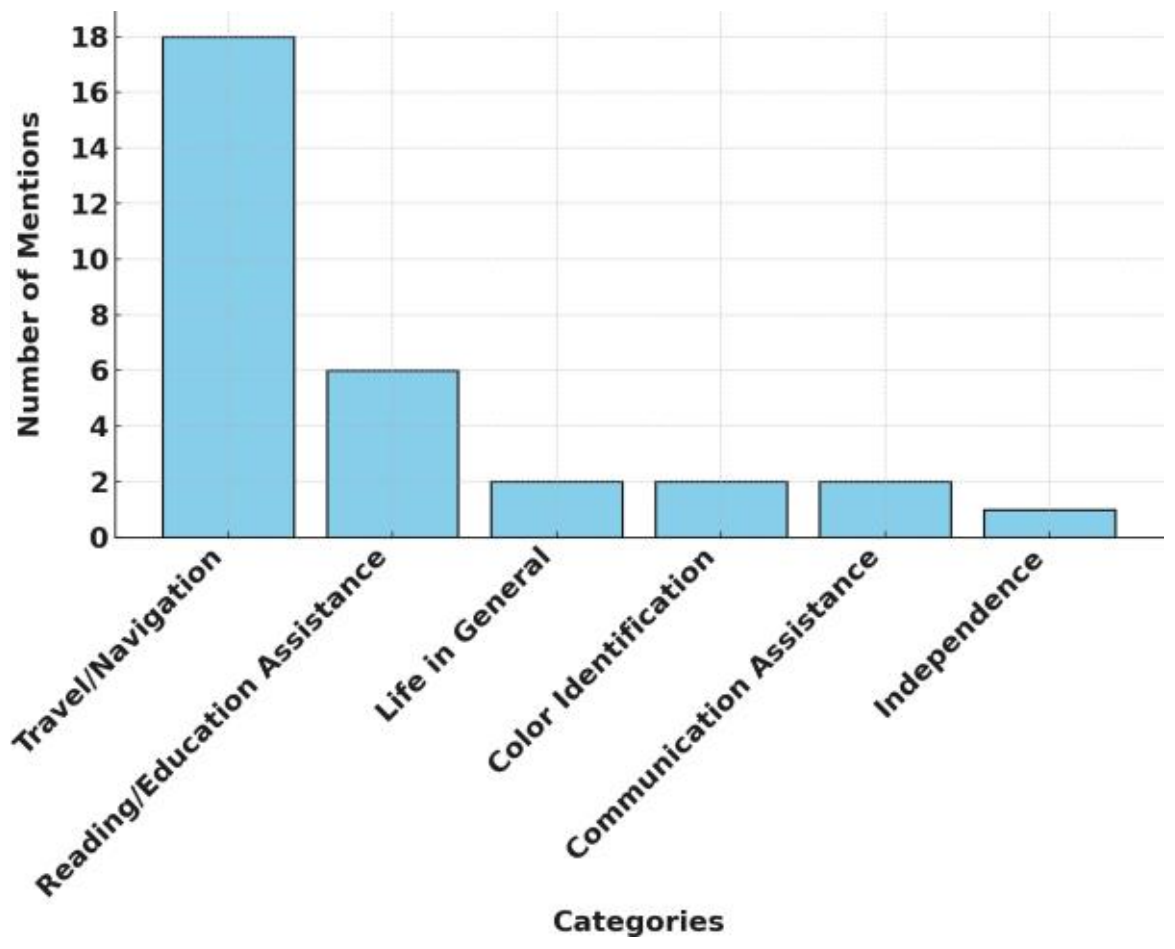

**Figure 4.** Summary of the responses to survey question Q7. Aspects of daily activities where assistive devices are most needed.

If so, please describe how and mention the type of technology.

Q7- For which aspect of your daily activities do you think you can most benefit from an assistive device?

**Responses to Q6**

In Q6, the focus shifts specifically to the use of technology as a means to assist with daily activities due to visual impairment, whereas Q2 broadly inquired about the smart devices participants use in general. As a result, a greater proportion of respondents (nearly half) now report using specialty assistive devices. However, smartphones remain the most frequently reported assistive device, highlighting their critical role not just as a mainstream technology used by both sighted and visually impaired individuals, but also as a powerful assistive tool for the visually impaired.

Notably, half of the participants still do not report using any specialty assistive devices, indicating that many rely primarily on general-purpose technologies such as smartphones and computers for accessibility. Another important observation is the wide variety of specialty assistive devices reported within a relatively small (9 out of 18) population. This suggests that there is no single dominant assistive device that has been able to address a need and is universally preferred among visually impaired users. The reported technologies include refreshable Braille displays Kapperman et al. (2018), screen magnification tools Baldwin et al. (2020), ORCAM Waisbourd et al. (2019) (a wearable AI-driven assistive device) , PenFriend Kendrick (2011) labelling system (audio labeling device), Victor Reader Trek Deverell et al. (2020) (a GPS and media player for blind users), eReaders, white canes, and Envision glasses Gamage (2024) (AI-powered smart glasses for visually impaired users).

**Responses to Q7**

The responses to this question were open-ended. The common themes from the responses are summarized in Fig. 4. The responses indicate that travel and navigation is the most frequently cited aspect of daily life where assistive devices are needed, with 18 mentions, highlighting the significant mobility challenges faced by individuals with visual impairment. With 6 mentions, reading and education assistance follows as the second category, emphasizing the need for tools that facilitate access to written materials and learning resources.

## *Navigational Challenges and Related Assistive Needs*

This section explores the challenges participants face when navigating their environment, both in immediate surroundings and over longer distances, as well as their confidence levels and areas where assistive devices could provide the most benefit.

Q8- What are your major challenges when navigating your surrounding environment?

Q9- How confident do you feel when navigating new places? (1-5, from (1) not confident at all to (5) very confident)

Q10- Do you have trouble navigating your immediate (short range) environment? An example is moving towards the front desk in a hotel or a bank after entering the building. Please explain the challenges you face in this case.

Q11- Do you have trouble navigating your long-range environment? An example is finding your way from home to your bank or favorite store. Please explain the challenges you face in this case.

Q12- Which aspects of navigating your surrounding environment can most benefit from an assistive device?

**Responses to Q8**

The responses to this question were open-ended. The key themes and categories in the responses are summarized in Fig. 5. These categories represent the most commonly reported challenges faced by IVI when navigating their surroundings.

Identified categories and examples:

1. Physical Barriers Description: Challenges related to obstacles, uneven surfaces, or structures in the environment. Examples: “Physical barriers, objects, higher than the level of my cane that I can run into.” “Curves and uneven surfaces, poor floor plan layout of unfamiliar surroundings.” “Bumping into people, walls, tripping.”

2. Orientation and Navigation Description: Difficulty in determining direction, navigating spaces, or locating target objects. Examples: "Orientation to how close or how far people or objects are from me." "Safe street crossings, finding entrances." "Using pathways inside parks to enter and exit. Finding a bench to sit on. Finding traffic signal pole buttons for crossing the street." "My greatest challenges when navigating my environment include streets and other items that are not very accessible. These include intersections with odd shapes, as an example, or intersections with opposite corners that are far more misaligned than normal."

3. Unfamiliar Environments Description: Struggles with navigating new or unknown areas. Examples: "My challenges when I am somewhere new and I am not familiar with an area." "I feel unsafe and unable to navigate in new environments and need a sighted person to accompany me."

4. Safety Concerns Description: Fears or risks posed by environmental hazards or lack of sensory feedback from modern vehicles or cyclists. Examples: "Finding transportation when distances are greater than walking permits, electric cars that do not make enough noise as well as speeding cyclists who do not make enough noise." "I feel unsafe and unable to navigate in new environments and need a sighted person to accompany me."

5. Social Barriers Description: Misunderstandings or lack of assistance from others in social situations. Examples: "Mostly others misunderstanding of me and how I get around as a blind person." "People are reluctant to lend arm/shoulder and don't understand that I request an arm to guide me."

6. Technology Limitations Description: Challenges with unreliable or inaccessible technology in certain environments. Examples: "It is challenging if I am in an unfamiliar place where it is hard to access WiFi, use GPS, reach anything with an iPhone." "Never learned GPS aids (and am desperate to do so)."

7. Auditory Limitations Description: Difficulty identifying or localizing sounds. Examples: "If I hear something, I cannot determine what or where it is."

8. Reliance on Memory and Planning Description: Dependence on memory or extensive planning for navigation. Examples: "I forget the roads and so use Google Maps for navigation."

**Responses to Q9, Q10, and Q11**

Figures 6, 7, and 8 summarize the responses to questions 9, 10 and 11. It should be noted that given the open-ended nature of questions 10 and 11, the same respondent could refer to multiple identified challenges. Figures 9 and 10 combine these results and provide the average confidence of the participants for each category of the reported challenges. Interestingly, the plot of Fig. 9 indicates that for short-range navigation the individuals reporting auditory challenges exhibited relatively high confidence overall, even higher than those who could not think of any specific issues, suggesting that those who rely on auditory cues for navigation may generally perform well. As a result, for this group, auditory distractions become particularly problematic, as they interfere with an essential navigational aid. Participants who reported no significant issues take the second level on average confidence navigating their environments. Conversely, for those who find obstacle avoidance, and unfamiliar environments as challenging, were associated with lower confidence in general. Those who mentioned unfamiliar environments, consistently reported reduced confidence levels, reinforcing the idea that familiarity or existence of a prior mental image of the environment is a key factor in feeling secure. The plot in Fig. 10 shows that participants who identified complex environmental layouts as a challenge reported the highest confidence levels, while those who did not report any issues (no significant issues) had slightly lower confidence. This aligns with our earlier observation on auditory challenges in short-range navigation, suggesting that individuals who recognize nuanced challenges, which may not be immediately apparent to others, are often the most experienced in navigation.

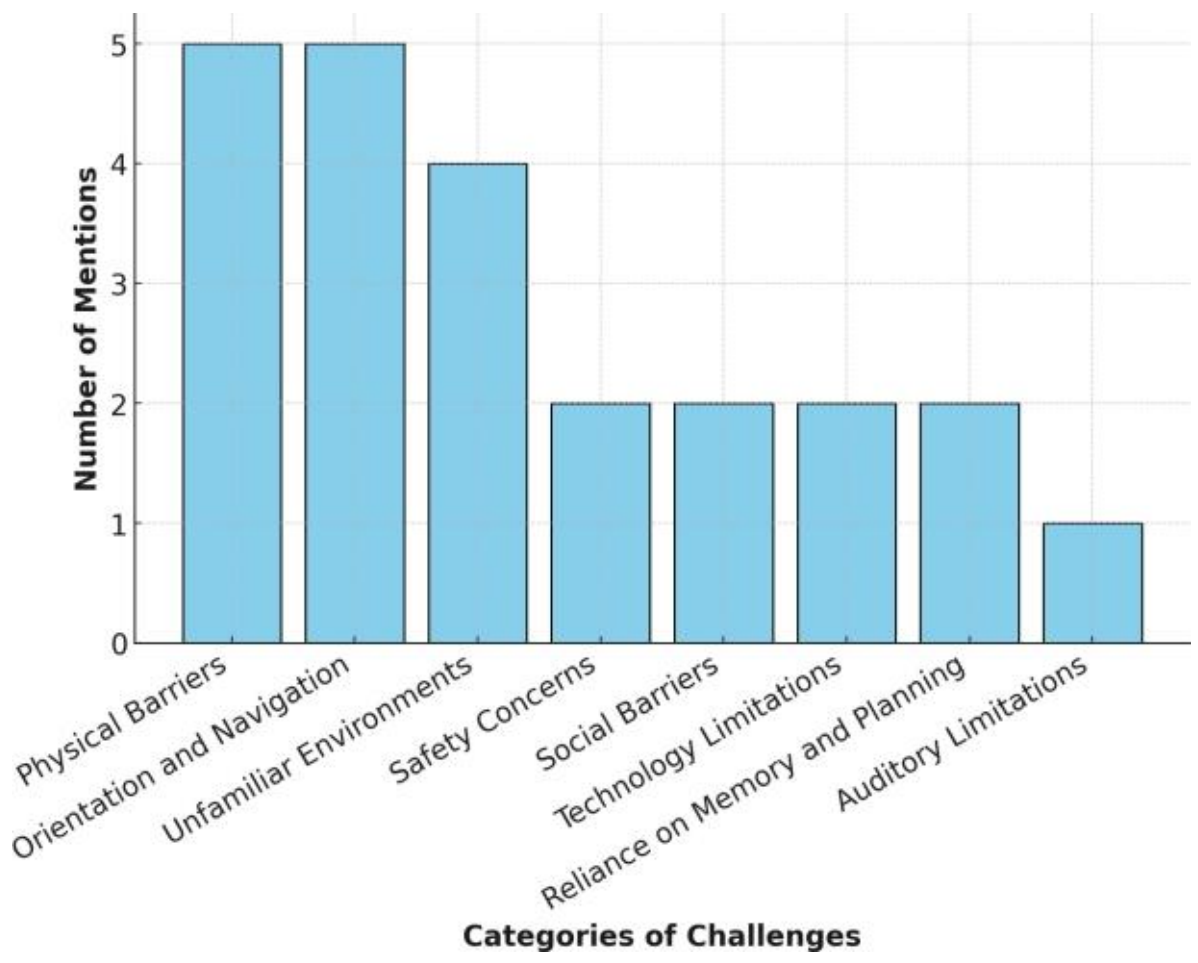

**Figure 5.** Summary of the responses to survey question Q8

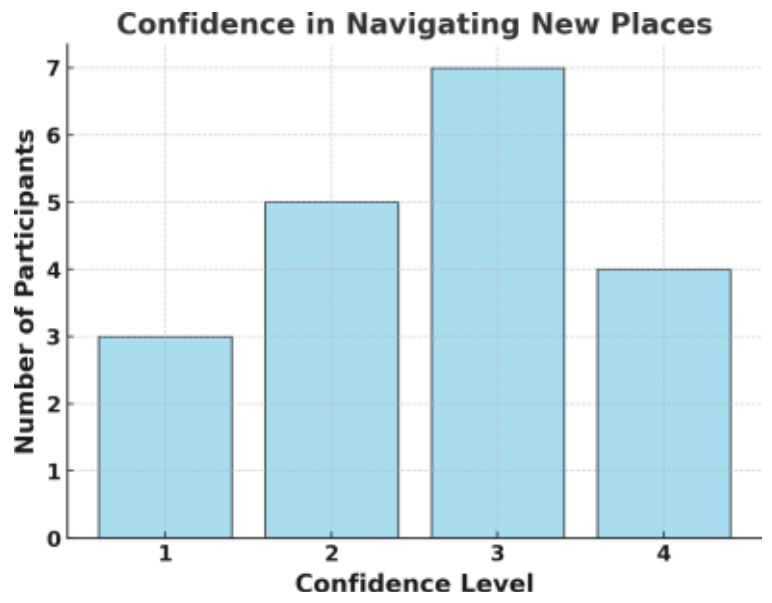

**Figure 6.** Summary of the responses to survey question Q9

**Responses to Q12**

The following categories are identified from the open-ended responses:

*Directional Assistance*: Tools to provide clear guidance towards specific destinations. Obstacle Avoidance: Detection and avoidance of physical barriers during navigation. *Final Feet Barriers*: Challenges in locating entrances or points of interest near a destination. *Unfamiliar Environments*: Navigating in places with no prior knowledge of layout or features. *Current Location Awareness*: Identifying one's position relative to surroundings. *Traffic and Signal Identification*: Interpreting traffic lights and street signs. *Navigating Around People*: Avoiding or maneuvering through crowded areas. *Identification of Objects of Interest*: Recognizing specific objects like traffic poles or benches.

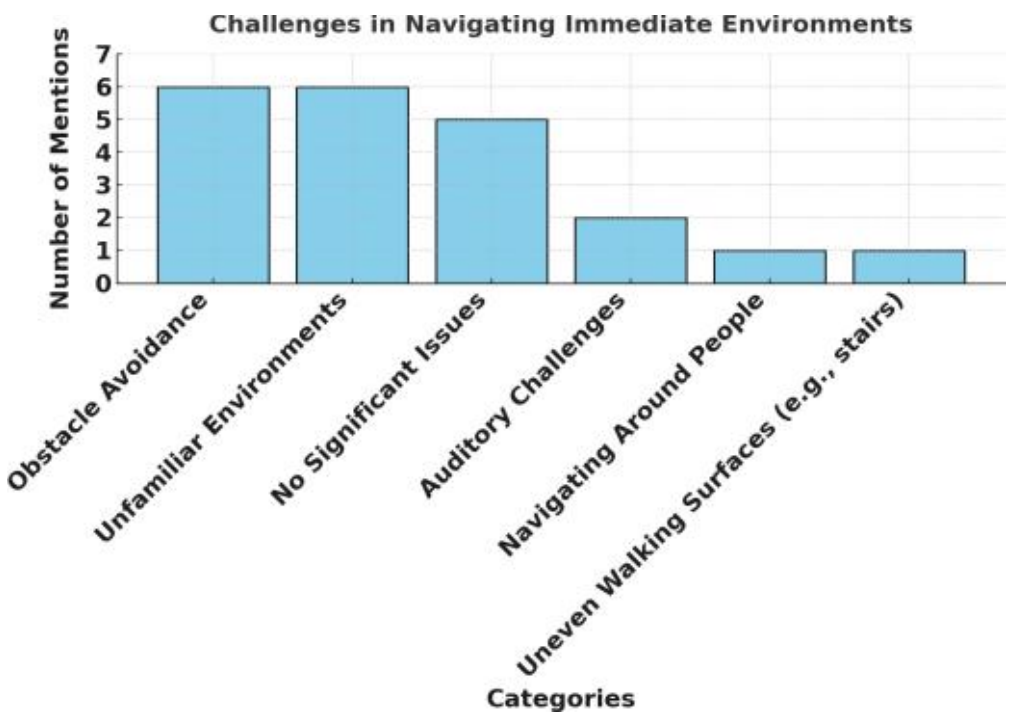

**Figure 7.** Summary of the responses to survey question Q10

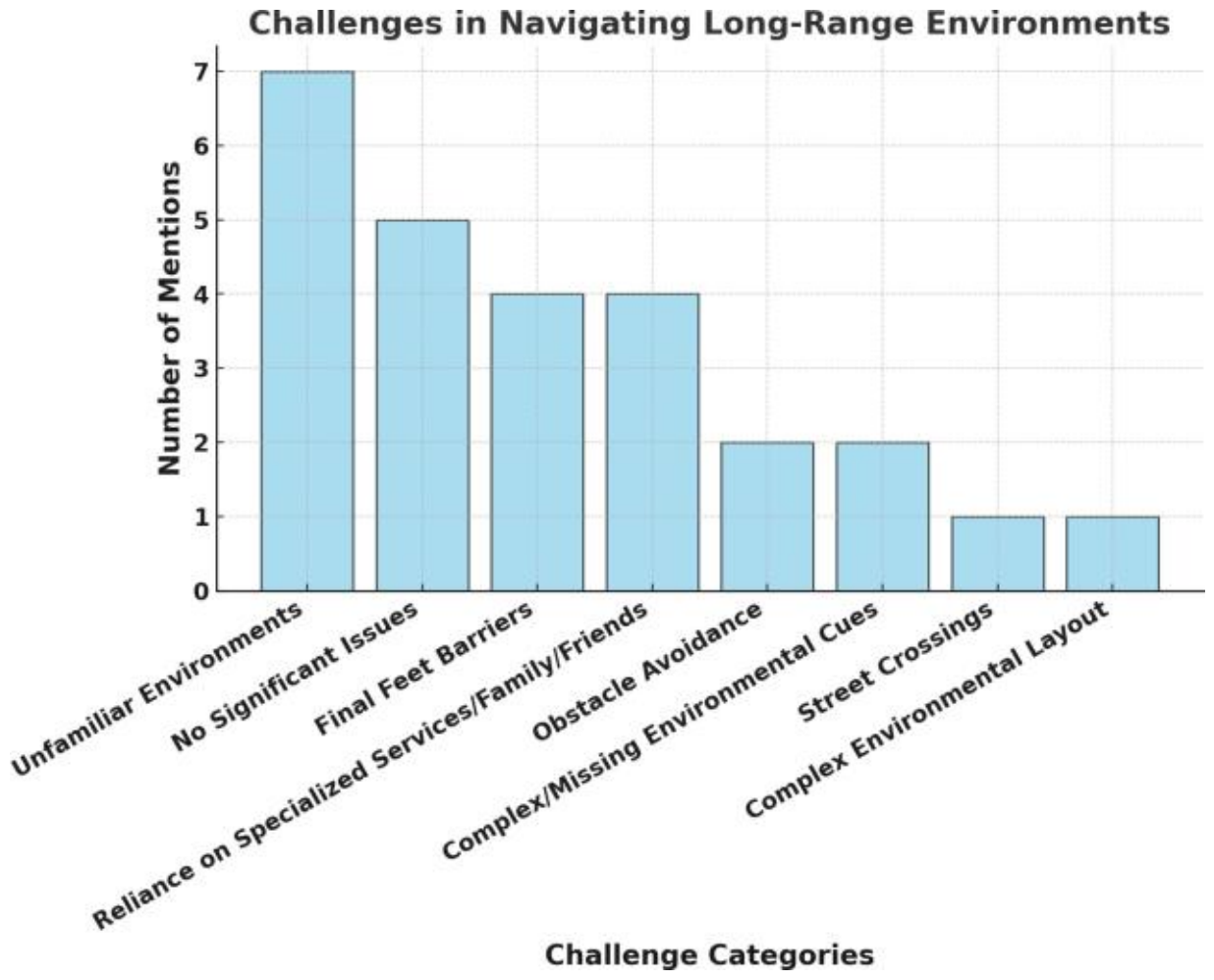

**Figure 8.** Summary of the responses to survey question Q11

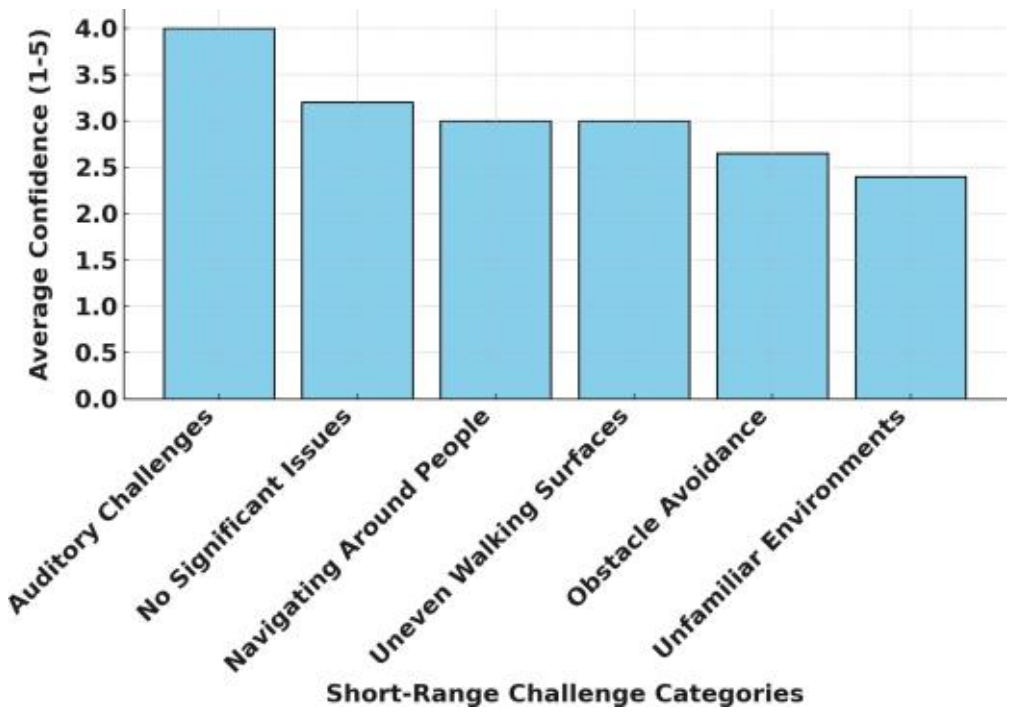

**Figure 9.** Average confidence for each short-range navigation challenge listed in Fig. 7.

*Street Navigation*: Following paths or roads while walking or traveling. *Navigation in Complex Layouts*: Managing routes with non-linear or unconventional layouts.

Figure 11 provides a summary of the results for the above categories. The responses highlight a wide range of focus areas, suggesting that navigation is a highly individualized experience, influenced by specific personal needs and environmental contexts. As such, different individuals see value in assistive devices for various purposes. Rather than pointing to a single, universal need, the data implies that a flexible or modular approach to assistive device design could be beneficial, allowing individuals to tailor solutions to their specific needs in a given environment or circumstance.

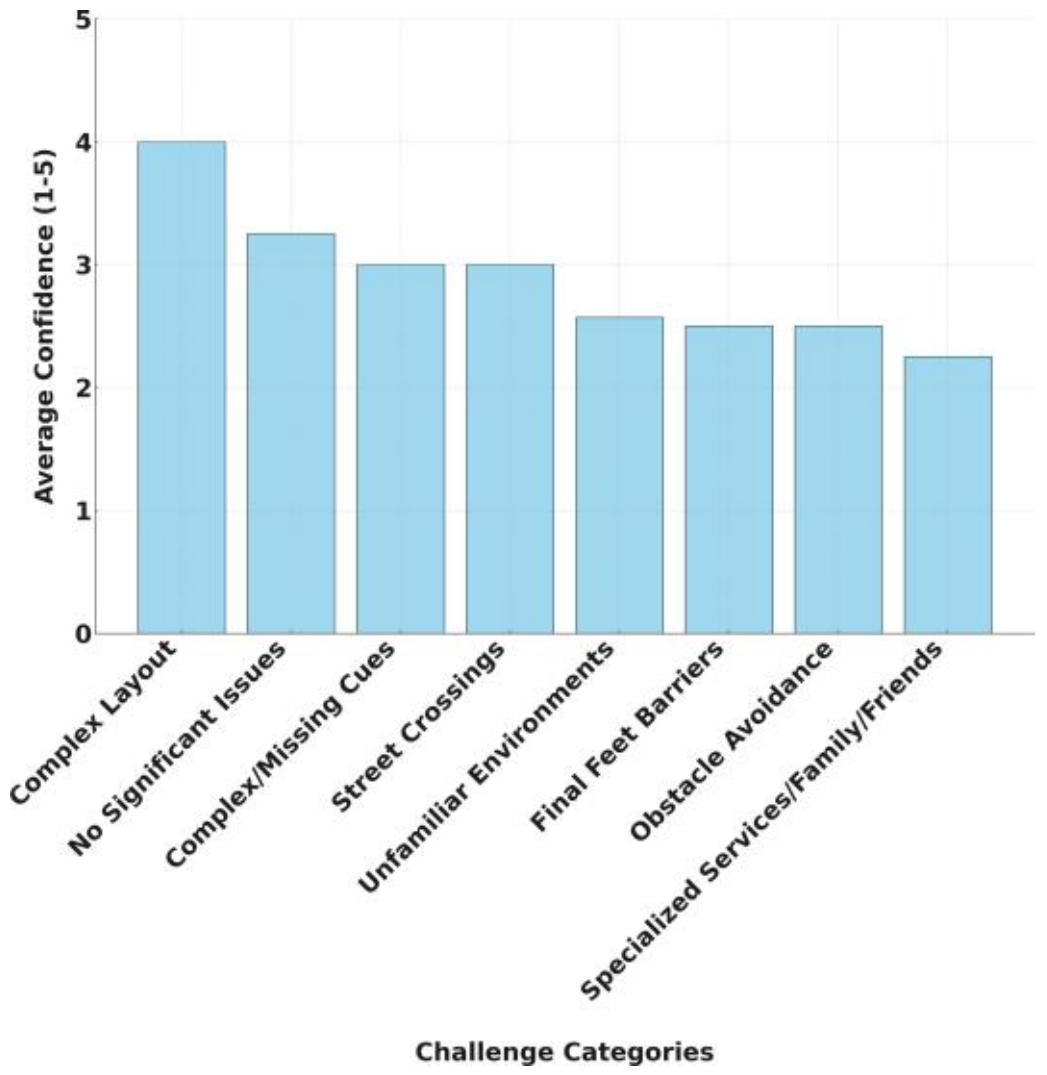

**Figure 10.** Average confidence for each long-range navigation challenge listed in Fig. 8.

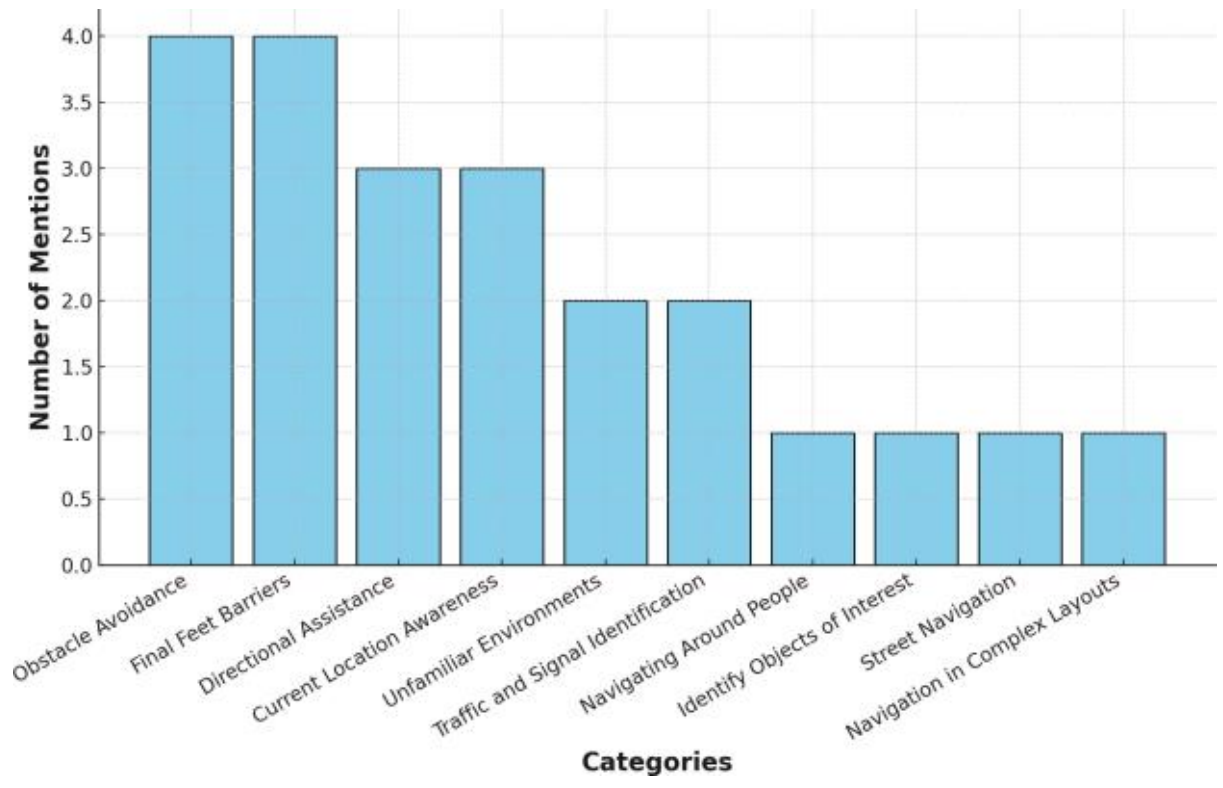

**Figure 11.** Summary of the responses to survey question Q12.

## *Traditional Assistive Navigation*

This section examines participants' experiences with traditional navigation aids such as white canes, guide dogs, and GPS devices, assessing their satisfaction levels and identifying limitations that non-traditional technologies should address.

Q13- Have you ever used "old technology" navigational aids such as white cane, guide dog, GPS, etc.? If yes, what did you use and how often?

Q14- How satisfied are you with the "Old Technology" devices you tried? (1-5, from (1) not satisfied at all to (5) very satisfied)

Q15- What are the shortcomings of the "Old Technology" navigation aids that should be addressed in the new technologies?

**Responses to Q13, Q14 and Q15**

As indicated in Figs. 12, 13, and table 1 white cane is the most commonly mentioned "old technology" navigational aid, followed by GPS. guide dog and combined audio/haptic were mentioned less frequently. In terms of satisfaction, guide dog received the highest average rating (4.67), while combined audio/haptic had the lowest (3.5).

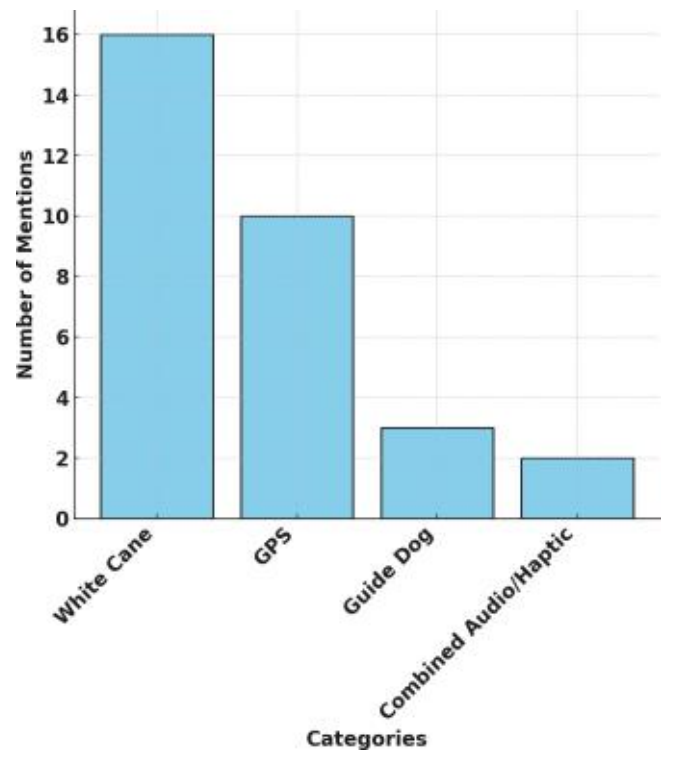

**Figure 12.** Summary of the responses to survey question Q13.

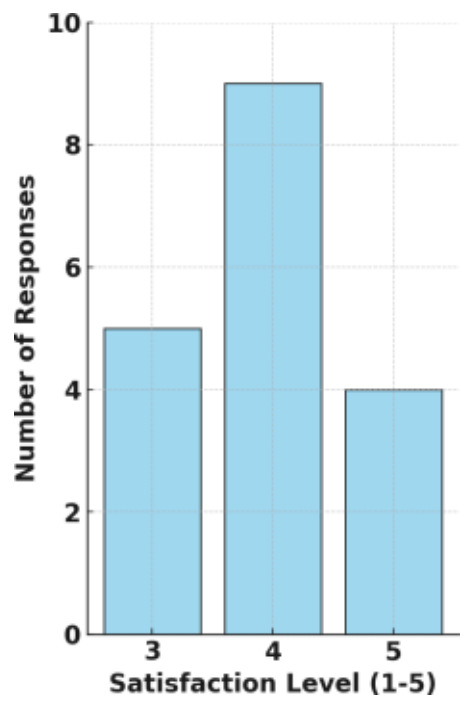

**Figure 13.** Summary of the responses to survey question Q14.

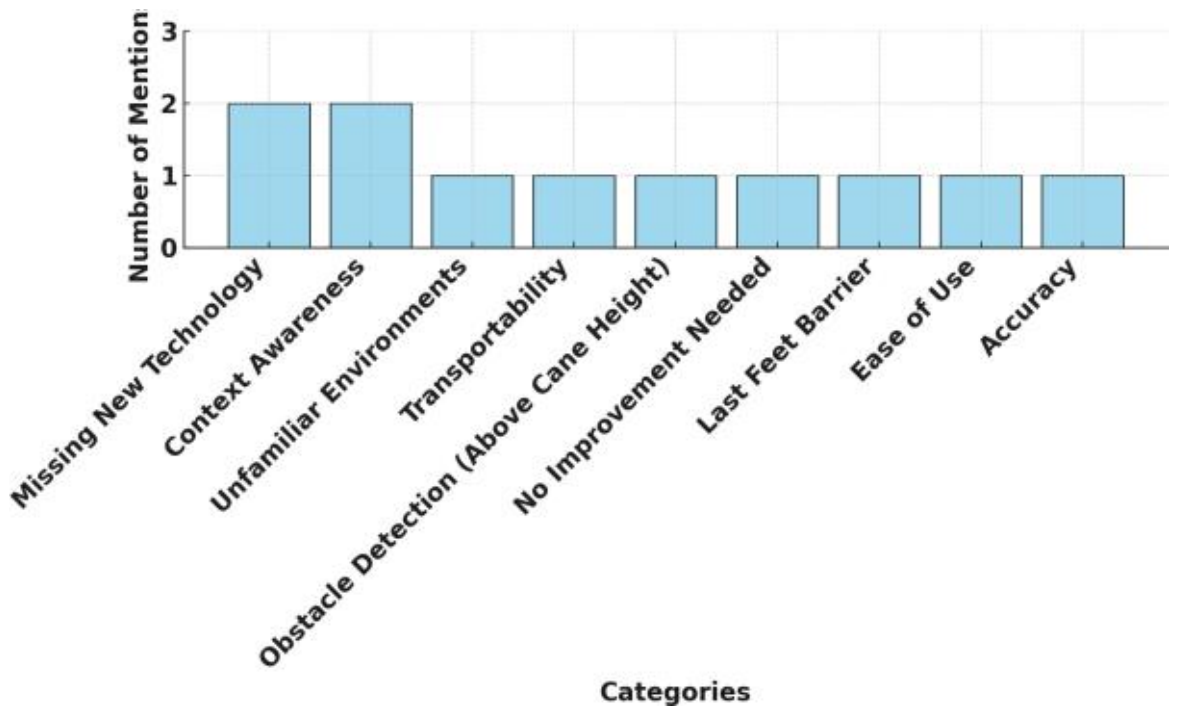

**Figure 14.** Summary of the responses to survey question Q15. Note: 8 out of 19 did not respond to this question.

As show in Fig. 14, When analyzing shortcomings of existing technologies, "Missing New Technology" and "Context Awareness" were the most frequently cited concerns, suggesting that users see a need to enhance old technologies using new technologies and also seek better adaptability to dynamic environments. Other concerns, such as obstacle detection (above cane height) and ease of use, highlight specific limitations that should be considered in new assistive technologies.

**Table 1.** Average satisfaction for mentioned old technology.

| | Ave Satisfaction | Mentions |
|---|---|---|
| White Cane | 4 | 16 |
| GPS | 3.8 | 10 |
| Guide Dog | 4.7 | 3 |
| Audio/Haptic | 3.5 | 2 |

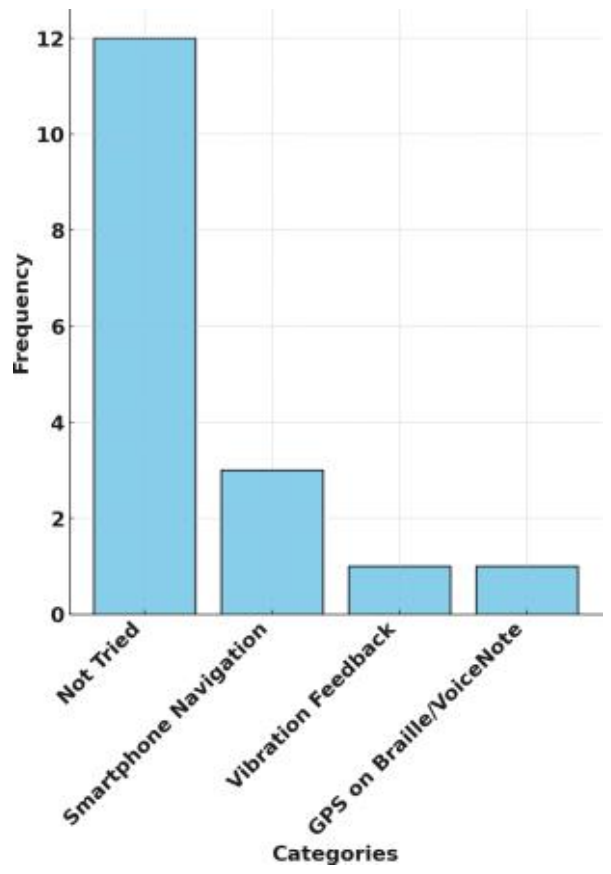

**Figure 15.** Summary of the responses to survey question Q16.

## *Non-traditional Assistive Navigation Technologies*

This section examines participants' experiences with newer assistive navigation technologies, evaluating their satisfaction, continued use, and reasons for adoption or abandonment.

Q16- Have you ever tried a "new technology" navigational aid for example one that is capable of providing real-time feedback about obstacles and can guide you towards your desired destination? If yes, please provide more details about the device.

Q17- How satisfied are you with the "New Technology" devices you tried? (1-5 from (1) not satisfied at all to (5) very satisfied)

Q18- Did you abandon the "New Technology" device you tried? If so why? If you kept using it please explain what you liked most about it.

Q19- Have you ever received training on how to use "New Technology" navigational aid devices (other than white cane, service dogs)? Yes/No, if yes, how long was the training. Do you remember the name of the product you received training for?

Q20- Based on your experience with navigation aid devices, do you think using such systems expand your daily activities and travel distance? (1-5, from (1) not at all, to (5) extremely)

**Responses to Q16, Q17, Q18**

According to Fig. 15, three respondents to Question 16 identified smartphone-based navigation tools as a "new technology". However, given that all 19 respondents reported using smartphones in response to question 2, it is reasonable to infer that most did not consider smartphone apps as a specialized new navigational aid technology for the visually impaired. If we exclude smartphones from the

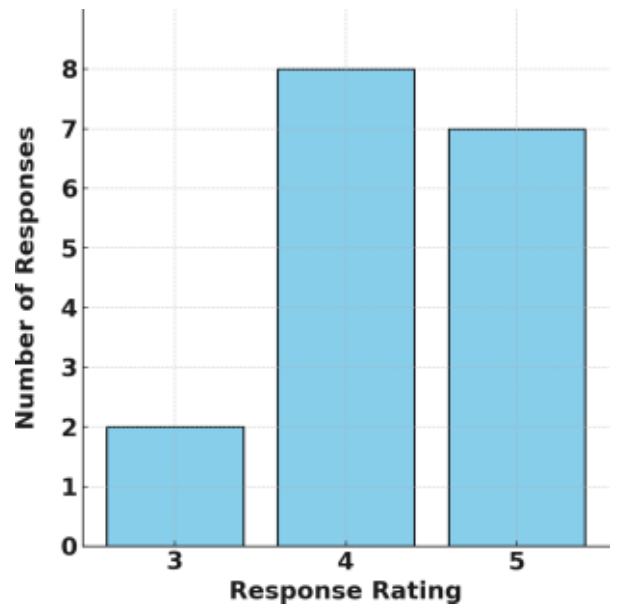

**Figure 16.** Summary of the responses to survey question 20.

definition of ”new assistive technology,” the data suggests that 15 respondents did not use any other navigational assistive technologies. This leaves only two respondents who have experience with such devices. Among them, only from the single participant who tried vibration feedback technology, we obtained a valid response to Q17 and Q18, where the respondent rated their satisfaction at 3 and cited discontinued access as the reason for abandonment. Despite extensive research and the development of numerous assistive navigation technologies Bakali El Mohamadi et al. (2024), their limited adoption by the visually impaired community highlights a critical gap that warrants further investigation into the factors preventing their success and widespread use. Figures

**Responses to Q19**

Out of the 19 individuals surveyed, 3 did not respond to this question. Among the remaining 16 individuals, only 3 indicated they had ever received training, where all were related to apps and smartphones.

**Responses to Q20**

Figure 16 summarizes the responses to this question, which reveal a notable gap between experience and perception regarding advanced navigational aids. While only two respondents had previously used new navigational aid technologies (Q16), under question 20 the majority rated their potential usefulness as high (4 or 5). This suggests that despite limited exposure or even prior negative experiences, respondents still recognize the value such technologies could offer if implemented effectively. Their responses indicate a belief that, while existing solutions may not have met their needs or were unavailable to them, the right technological approach could significantly improve their mobility and independence. This highlights the need for better outreach, improved usability, and affordability, ensuring that such assistive technologies are both accessible and designed to truly address the challenges faced by individuals with visual impairments.

### *Information and Feedback Preferences*

This section explores participants’ preferences for the type and format of feedback provided by assistive navigation devices, including the balance between direct guidance and spatial awareness.

Q21- When using a navigational aid device, do you prefer to have multiple types of feedback e.g. both audio and haptic (vibration for example)? please specify what type(s) of feedback is preferred and explain why.

Q22- How important is it for you to understand the arrangement of objects and the surrounding spatial setting when moving around a new indoor or outdoor environment? Do you think availability of such information helps you plan your route and better navigate the environment? Or do you only need to know in what direction to move (how to navigate)?

Q23- Do you prefer your navigation assistive device to give you navigational guidance (move left, right, straight, etc.) or do you prefer that it provides you with information about the surrounding (arrangement of obstacles) and let you decide about your route and movements.

**Responses to Q21**

As shown in Fig. 17 the respondents identified combinations of audio, tactile and haptic feedback, indicating that many find value in using complementary modalities for navigation aids. This underlines the importance of designing flexible navigation aids capable of supporting multiple modalities and customization to address diverse user needs effectively.

In summary, the responses include the following justifications for mentioned modalities. Audio feedback was justified for its clarity and the ability to provide detailed descriptions of the surroundings. Respondents who preferred audio alone found it sufficient and reliable, emphasizing its straightforward nature and ease of understanding in most situations. Haptic, including vibrations, was appreciated for its speed in delivering quick notifications. It was also seen as essential in scenarios where auditory cues might be less reliable, such as in noisy environments or for individuals with hearing impairments. This feedback mode was particularly valued for providing simple, immediate alerts. The combination of audio/haptic feedback was the most preferred category. Respondents highlighted how these two modalities complement each other, with audio offering detailed explanations, haptic feedback providing faster, more immediate notifications. This combination was also valued for its redundancy, ensuring critical information is not missed if one mode fails or is inaccessible due to environmental factors.

The strong preference for audio and haptic feedback among respondents may be influenced by their familiarity with smartphones, which commonly provide both vibration and auditory cues for interaction. Given that all 19 respondents reported using smartphones, while only 2 had experience with assistive technologies beyond smartphones, it is likely that prior exposure to these modalities played a role in shaping their preferences. This familiarity may have created a positive association with audio/haptic feedback, reinforcing their selection when asked about navigation aids. Additionally, respondents may have found it difficult to propose combinations of modalities they have never encountered, further contributing to the dominance of audio/haptic choices in the survey responses.

**Responses to Q22, Q23**

Figure 18 presents a summary of the responses, categorized using sentiment analysis assigning each response an importance level on a scale from 1 to 5. As summarized in this figure, the survey responses to question 22 highlight the importance of situational awareness to those with visual impairment. A significant majority of respondents emphasized the critical role of understanding the arrangement of objects and obstacles in their surroundings. Many noted that such information enhances their ability to plan routes effectively, avoid unexpected obstacles, particularly overhead hazards, and navigate with greater confidence. Some participants provided real-world examples, such as encountering parked vehicles in non-designated areas or changes in indoor layouts, to illustrate how spatial awareness could have prevented confusion or accidents.

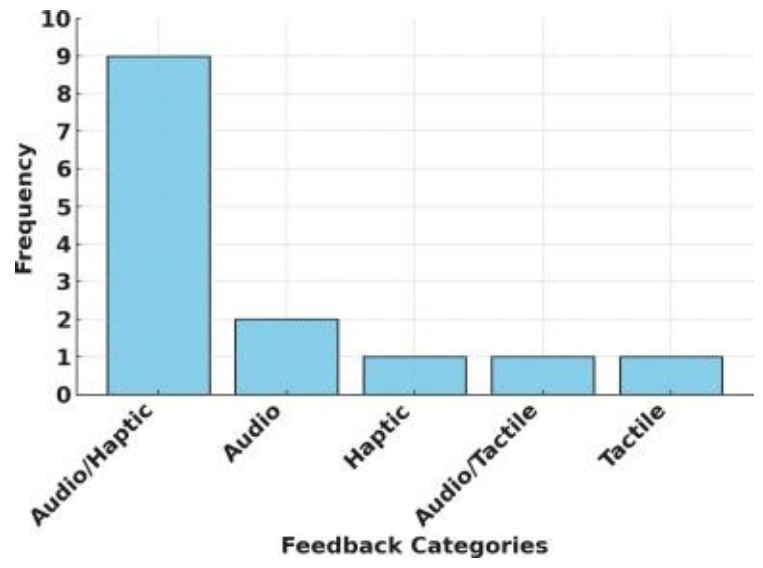

**Figure 17.** Summary of the responses to survey question Q21.

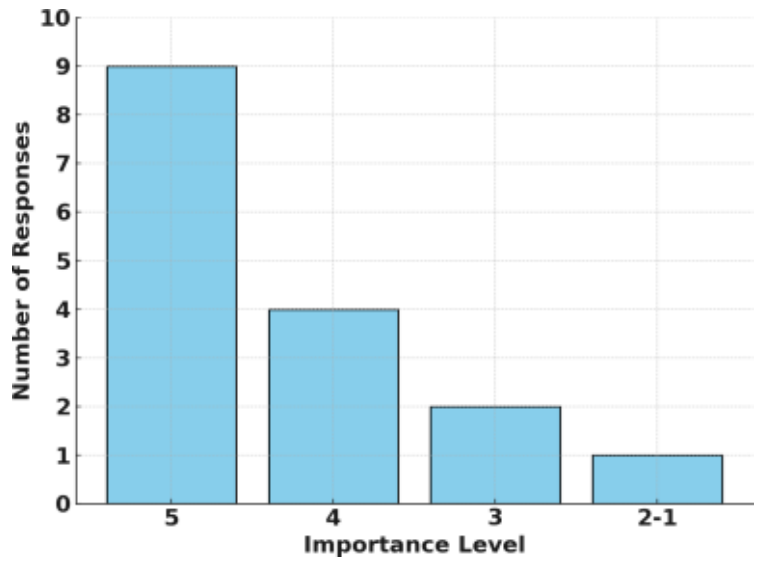

**Figure 18.** Summary of the responses to survey question Q22 per sentiment analysis results.

These responses are in line with those for questions 8 (Fig. 5), 10 (Fig. 7), and 11 (Fig. 8) which indicated that navigating unfamiliar indoor and outdoor environments is one of the greatest challenges faced by individuals with visual impairment. The objective of high-level spatial layout information may be to mimic the sense of familiarity, even in new environments, by providing users with a priori mental map of their surroundings.

While most respondents found spatial awareness essential, a smaller subset indicated that its usefulness depends on the context. Some preferred to receive only selective information, as excessive details could be overwhelming. A few respondents expressed a preference for a more minimalistic approach, stating that directional guidance alone (e.g., turn left, go straight) would suffice for their navigation needs.

Overall, the responses suggest that assistive navigation technologies should be designed with flexibility in mind, allowing users to access varying levels of spatial detail based on their personal preferences and situational demands. A system that integrates both directional guidance and environmental awareness may better support this goal.

While Q22 explored the overall importance of environmental awareness for navigation, Q23 was designed to further probe how individuals prioritize this information when presented with a choice between two distinct guidance methods: direct step-by-step navigation instructions versus environmental awareness for independent route planning. By framing the question as a decision between these two approaches, we aimed to gain a more nuanced understanding of how users perceive their relative benefits and limitations.

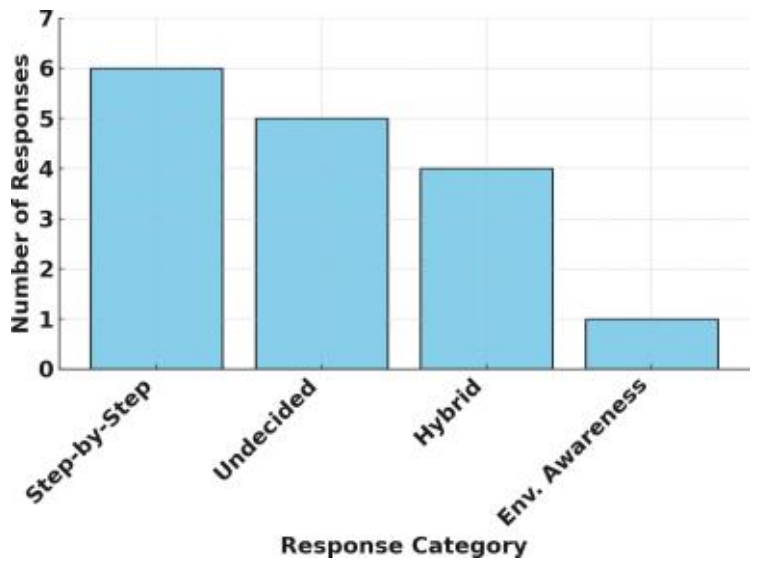

**Figure 19.** Summary of the responses to survey question 23.

In this case, a significant portion of respondents preferred step-by-step navigation guidance, where the assistive device provides explicit directional instructions such as ”turn left” or ”go straight.” These individuals emphasized the importance of clear and reliable guidance, noting that precise directions would help them move efficiently and confidently through an environment. Only one out of 16 valid responses expressed a preference for environmental awareness rather than direct navigation commands.

Notably, despite the question not explicitly offering a hybrid option, a third category of respondents favored a hybrid approach, stating that both directional guidance and environmental awareness could be useful depending on the context. Some suggested that an ideal system should allow users to switch between modes based on their needs, with step-by-step navigation being more helpful in structured environments like streets, while spatial awareness is more beneficial in open or complex spaces such as parks or indoor settings.

Finally, several respondents were undecided, indicating that they would need to try both methods before determining which works best for them. Figure 19 summarizes the responses to question 23.

### *Preferred Integration Platform: Smart Devices vs. Traditional Aids*

This section examines participants’ preferences for integrating assistive navigation technology into either smart devices or traditional aids, such as white canes, and the practicality of each option.

Q24- Assume a navigation aid device is incorporated into your smartphone: When navigating, would you feel comfortable using one hand to hold an object like a smartphone to receive information about the environment? Please explain your rationale.

Q25- Assume a navigation aid device is incorporated into your old technology assistive device, for example incorporated into the handle of your white cane: Do you feel comfortable using such a device? Do you prefer this over the case where the technology is incorporated into your smartphone?

**Responses to Q24**

Given the widespread use of smartphones, we explored whether greater efforts should be made to integrate navigation aids into these devices, including hardware accessories in addition to software, to enhance adoption and usability. The familiarity and accessibility of smartphones may provide a natural platform for assistive technologies, reducing the learning curve and increasing the likelihood of widespread acceptance. Understanding their perspectives on this trade-off between convenience and potential physical limitations can help researchers determine whether smartphone-based solutions warrant further investigation.

Of responses, the ”Yes” category represents 12 responses (63%), and the ”No” category represents 7 responses (37%). The ”Yes” group expressed comfort with using one hand to hold a navigation device, citing familiarity with smartphones as a key factor, as many already use them for similar purposes like audio feedback or navigation apps. They emphasized adaptability, with some highlighting the importance of accommodating other tools like canes or preferring hands-free options in certain scenarios. Several respondents viewed it as a skill they could learn or adapt to, while others mentioned that holding a phone for navigation aligns with their daily habits, making it practical and manageable. Despite conditional preferences for hands-free solutions, the overall sentiment was that using a smartphone in this way is feasible and aligns well with their current practices or experiences.

The ”No” group expressed discomfort with using one hand to hold a navigation device, primarily due to a preference for hands-free solutions such as smartwatches or wearable devices. Some emphasized the need to keep their hands free for other tools, like a cane or guide dog leash, or to ensure safety and practicality while navigating. Concerns about potential challenges, such as dropping or losing the phone, fumbling during use, or theft in public spaces, were also mentioned. Overall, the respondents in this group felt that hands-free options would provide greater convenience, security, and efficiency in navigating their environments.

The responses suggest that while a majority of participants are comfortable using one hand to hold a smartphone or navigation device, the feedback highlights the need for flexibility in designing assistive technologies. Given prior preferences for audio/haptic feedback, it is important to interpret this as an opportunity to explore augmentative technologies, such as tactile displays, while ensuring that audio and simple haptic feedback (e.g., vibrations) remain central to the design. These feedback modalities should cater to those who may not feel comfortable with one hand fully occupied, providing accessibility and usability across diverse preferences. Additionally, while some respondents in the ”No” group expressed discomfort with holding a device in specific scenarios, it is likely they might adapt in other circumstances, suggesting the need for technologies that support varying use cases and preferences.

**Responses to Q25**

This question stems from recent efforts to enhance assistive technologies by building upon existing solutions, such as smart canes Saaid et al. (2016), to improve their functionality and performance. Of the 19 participants, 6 did not respond to this question. Among the responses, there were 7 ’No’ (including 4 conditional) and 6 ’Yes’ (including 2 conditional).

For those who responded ”No”, the primary reasoning centered around a preference for alternative solutions over modifying the white cane. Some explicitly favored using a smartphone instead, while others rejected the idea without providing further justification. Among those who responded ”Yes”, the reasoning was more straightforward, with respondents expressing direct acceptance of integrating technology into the cane without raising concerns or conditions.

Respondents who gave a ”Conditional No” highlighted concerns about integrating technology into white canes, specifically citing issues with reliability, weight, and functionality. Many worried that the added technology might malfunction or fail, which could compromise the cane’s primary purpose as a dependable mobility tool. The increased weight of the cane was another frequent concern, as it could lead to strain or discomfort during use, particularly for those with prior physical limitations or extended use scenarios. Additionally, examples of existing products like the WeWalk WeWALK (2025) cane were cited as evidence of limited utility, with some respondents feeling that such devices do not provide sufficient coverage or justify the added complexity. These reservations underscore the importance of maintaining practicality and simplicity.

Some respondents who were conditionally open to the idea of integrating navigation aids into white canes expressed their willingness with specific reservations. Many highlighted the importance of ensuring the device remains lightweight and durable, as a heavier or fragile cane could hinder its usability and reliability. Compatibility with existing feedback mechanisms was another major concern, particularly with vibrations potentially interfering with the environmental feedback through the cane that users rely on to navigate effectively.

## *Financial Considerations*

This section explores participants’ current spending on navigational aid devices and their willingness to invest in such technologies.

Q26- On average, how much money do you spend annually on obtaining or using navigational aid devices?

Q27- How much money are you willing to spend annually on obtaining or using navigational aid devices?

**Responses to Q26**

A few respondents included the cost of buying a smartphone or transportation expenses (such as paratransit) in their responses. However, since all participants are already smartphone users, we excluded this cost to focus specifically on navigational aid devices beyond smartphones. Similarly, transportation costs are a common necessity for all participants and, as a service rather than a device, they do not fit within the scope of this analysis. Therefore, we also removed these expenses.

To ensure a fair and meaningful comparison, we carefully examined each response and extracted only relevant costs associated with dedicated assistive technologies, such as cane replacements, screen readers, braille displays, and other specialized assistive devices like the Victor Reader Trek. For respondents who reported multi-year purchases, we annualized the cost by dividing the total expense over its estimated lifespan. After refining the responses, we grouped them into cost brackets. The results are presented in Fig. 20.

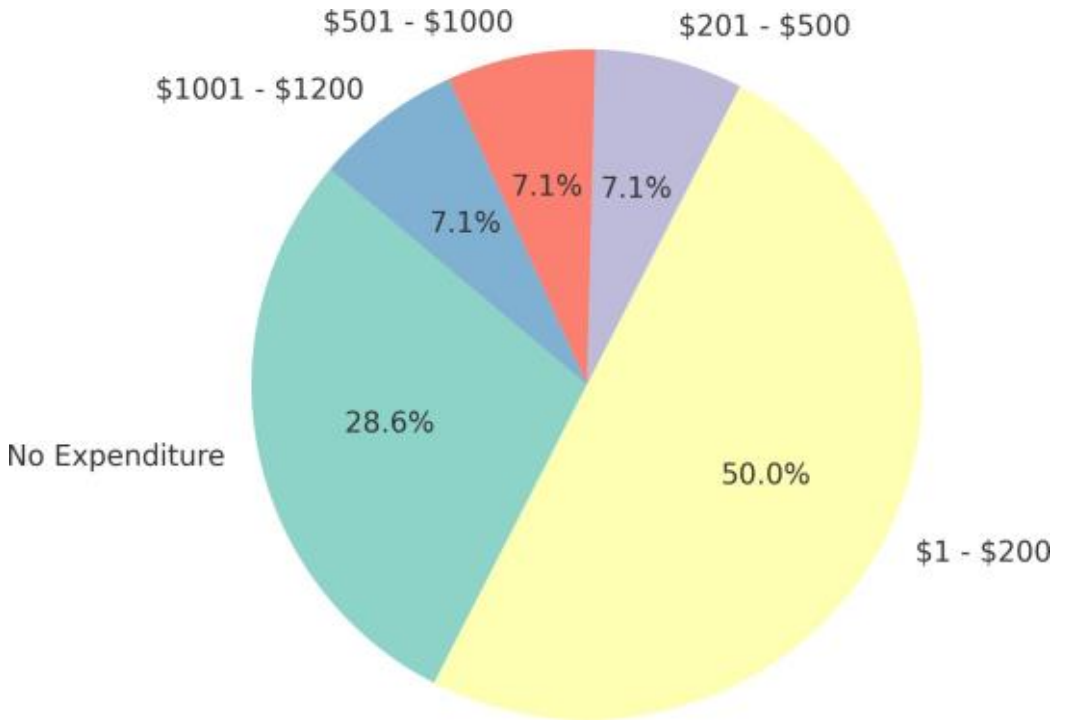

**Figure 20.** Summary of the responses to survey question 26.

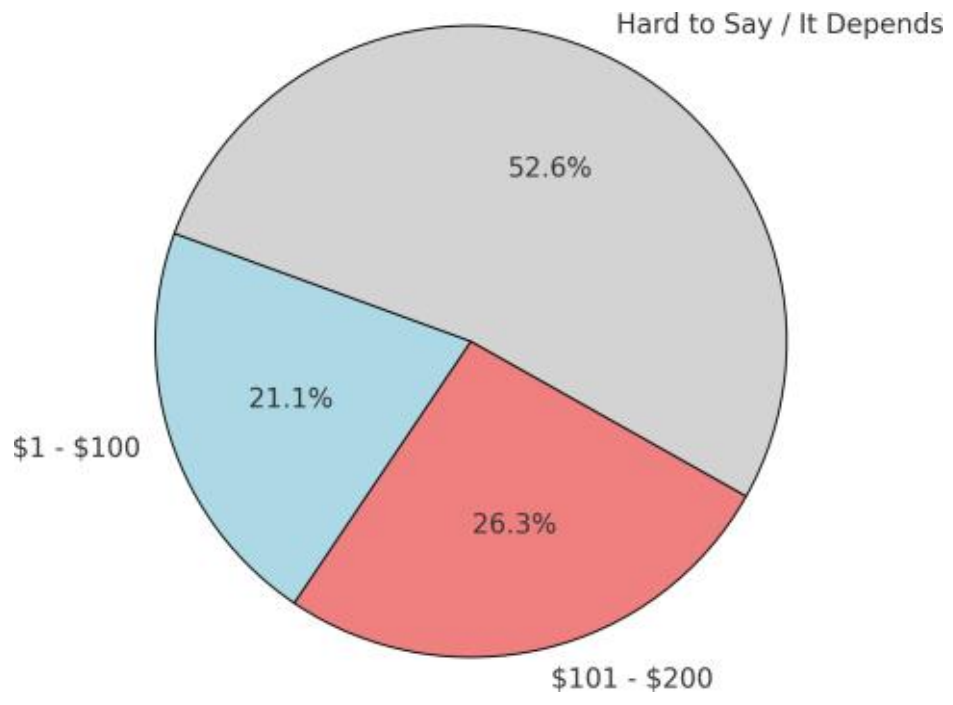

**Figure 21.** Summary of the responses to survey question 27

The results show that a significant portion of respondents spend little or nothing on dedicated navigational aid devices, relying instead on free smartphone apps and general-purpose tools. A moderate number of participants incur low to mid-range expenses, primarily for periodic cane replacements or occasional purchases of assistive technology. A small group invests relatively higher amounts in specialized devices such as screen readers, or braille displays. These results may reflect the financial constraints often faced by individuals with visual impairment Wickramaarachchi et al. (2023), emphasizing that cost remains a major barrier to the widespread adoption of assistive technologies. Therefore, affordability should be a primary consideration in the development of new assistive devices. However, the significantly inflated costs of blindness-related technologies have often posed a major barrier to their adoption. This point is poignantly captured by part of one participant's response to question 16: "To be honest with whoever is reading this, technology for the blind is horrendously expensive simply because not many people use them. It's just the kind of world we live in. So I probably wouldn't be able to afford it but I'm definitely open to new ideas and technologies to make the world a better and more accessible place."

**Responses to Q27**

Figure 21 summarizes the responses to question 27 indicating that half of the participants are willing to invest less than $200 annually on navigational aid devices. This suggests that IVI either lack the financial capacity to spend more or, based on experiences, do not expect a significant return on their investment in such technologies.

However, the majority of respondents (52.6%) were unsure or stated that their spending would depend on various factors, refusing to specify an exact amount. This might imply that if a truly transformative assistive technology were introduced, one that significantly enhances their quality of life, many might be willing to stretch their financial resources to afford it. Their response of "Hard to Say / It Depends" could also reflect a general hesitation to mention a specific number due to common financial limitations within the community of IVI.

These findings again highlight the critical role of cost in the adoption of assistive technologies. Developers must recognize that affordability is a determining factor for widespread accessibility and adoption. Creating cost-effective solutions with clear, demonstrable benefits will be essential in ensuring that assistive navigation technologies reach those who need them most.

## *Workshop Highlights*

In the following, we present a detailed analysis of the workshop discussions for each of the four themes described earlier in section .

## *Navigation Challenges*

a) **Physical Obstacles and Environmental challenges:** Participants frequently reported challenges with physical obstacles in their environments. In particular, obstacles above waist level—such as tree branches, protruding structures, vehicles partially obstructing sidewalks, and hanging signs or banners, often go undetected by canes, creating hazardous navigation conditions. Unexpected obstacles such as bicycles or improperly parked scooters and uneven surfaces further complicate safe passage. Crowded areas also pose significant challenges. The overlapping noises in such spaces can be disorienting, and navigating around people who may not yield space increases the risk of collisions. Additionally environments like parks and areas with irregular obstructions such as awning supports or overgrown shrubbery further complicate free movement and orientation. The reduced auditory cues from electric or hybrid vehicles coupled with intrusive background noises like jackhammers or leaf blowers were mentioned, making safe navigation difficult. It was noted that noise in general can overwhelm other sensory cues. Finally, navigating unfamiliar environments, such as offices or stores, can be particularly daunting due to the lack of clear orientation aids and indistinguishable layouts. While home environments tend to be more navigable due to familiarity, moving to a new residence or community requires relearning layouts, which can be time-consuming and effort-intensive.

b) **Inadequate Infrastructure:** The discussion highlighted signnificant concerns regarding the lack of appropriate infrastructure in areas like parking lots. These spaces often fail to provide clear structural or haptic cues and designated pathways for pedestrians with visual , severely limiting independent navigation and sometimes even posing safety risks. Additionally, inconsistent layouts in parks or large open areas, particularly when paths are winding or dead-ends, add to the challenges.

c) **Limitations of Assistive Devices:** The limitations of white canes, which do not protect against waist-up obstacles and overhead hazards, and the unavailability, complexity or inaccessibility of latest technologies were highlighted as a significant gap in current assistive technology.

d) **Impact on Independence and Mobility:** These navigational challenges impact the independence and confidence of IVI, affecting their willingness to engage in both routine and new activities outside their homes.

### *Type of Information Needed to Better Navigate the Surroundings*

a) **Obstacle Awareness and Detection:** Participants consistently highlighted the need for tools to provide information about static and dynamic obstacles in their environment. This includes barriers above waist height, uneven surfaces, curbs, vehicles parked on sidewalks, and electric cars, which are silent and difficult to perceive.

b) **Environmental Mapping and Layout Information:** Participants stressed the importance of having access to environmental layouts. They noted that knowing the spatial arrangement, from room configurations like the placement of furniture in buildings to the external structure of parks, gated communities, or parking lots, can significantly aid navigation. In particular, layouts that account for non-intuitive elements, such as winding paths or irregularly organized areas, are critical. Participants also expressed interest in tools that help improve mental visualization of these environments, which could further support independent navigation and enhance safety.

c) **Directional and Pathway Guidance:** Clear, step-by-step directional guidance was identified as crucial for confident navigation. Participants underscored the importance of receiving detailed instructions when entering unfamiliar buildings or outdoor spaces. This guidance should include information on how to locate key features such as entrances, exits, or desks, and provide clear directions on traversing paths while avoiding hazards like construction areas or parked cars. Such systematic and explicit navigation aids are seen as crucial for enhancing both safety and independence.

d) **Real-time Feedback:** Real-time updates about changes in the environment or the presence of dynamic obstacles were highlighted as vital for safe and efficient navigation. Participants emphasized that unexpected changes, such as temporary barriers, shifting obstacles, or queues in stores, can disrupt a planned route. By receiving live information on such changes individuals are better equipped to adapt their navigation strategies.

e) **Street Navigation and Safety Features:** For outdoor navigation, participants noted the need for rich contextual details about their immediate surroundings such as information about stairs, curbs, or uneven sidewalks. Participants also emphasized the importance of indicators at crosswalks, information about slopes or ramps, and assistance in determining when it is safe to cross streets, especially in areas with electric or hybrid vehicles that often operate without generating much noise.

f) **Location of Key Elements for Indoor Navigation:** Participants emphasized that having detailed information about key elements, like the locations of queues, desks, or counters, can make navigating unfamiliar indoor spaces easier. Additionally, in large indoor spaces, importance of knowing the exact placement of specific rooms or exits, was noted.

### *The Preferred Type of Feedback From an Assistive Device”*

a) **Audio Feedback:** Audio feedback emerged as highly favored among participants due to its ease of use and hands-free and user-friendly nature. It effectively delivers real-time assistance through turn-by-turn directions, obstacle warnings, and providing spatial layout details. Participants found audio cues especially useful for identifying key features like bus stops, intersections, and pathways in both indoor and outdoor environments. By providing clear and actionable instructions, audio feedback plays a critical role in supporting independence and safe navigation.

However, several challenges were identified. Background noise in busy environments, such as urban streets or crowded restaurants, can make it difficult to hear instructions. There were also safety concerns about audio feedback, particularly when using earbuds, that might block critical ambient sounds, like those of approaching vehicles which are crucial for safety. The preference is for audio feedback that does not mask these environmental sounds to maintain situational awareness, such as approaching vehicles or other auditory cues in the environment. To address these issues, participants suggested solutions like bone conduction headphones, wearable speakers positioned near the ears, and dynamic volume adjustments tailored to different environments. Additionally, participants emphasized the need for customizable verbosity levels to avoid information overload.

b) **Haptic Feedback:** Haptic feedback was regarded as a reliable complement to audio, particularly in scenarios where auditory cues may be ineffective. Vibratory cues were seen as useful for obstacle detection, with participants suggesting varying intensity or frequency to indicate proximity or urgency. Haptic signals were also considered effective for guiding users along predefined paths, offering a discreet and non-intrusive way to stay on course, especially in dynamic environments like crowded streets or public transport hubs.

Participants favored wearable designs, such as wristbands or watches, to integrate haptic feedback seamlessly into their routines. This approach reduces reliance on handheld devices, making it easier to navigate while carrying other items. For real-time navigation, participants stressed the importance of haptic systems that can adapt quickly to environmental changes, such as sudden obstacles or construction zones. The immediacy and precision of haptic feedback make it an invaluable addition to navigation technologies.

c) **Tactile Displays:** Tactile displays, inspired by Braille, were appreciated for their potential in pre-navigation tasks and detailed spatial exploration. These displays allow users to create mental maps of their environment, which is particularly helpful for navigating indoor spaces like airports, restaurants, or banks. By providing a bird's-eye view of obstacles, seating arrangements in indoors public settings, and key landmarks, tactile displays could enable users to understand the layout of a space before moving through it.

Despite their potential, tactile displays come with challenges. Non-Braille users expressed concerns about the learning curve associated with reading tactile patterns. Slow interpretation of tactile information was seen as a limitation for real-time navigation, although simplifying tactile patterns into intuitive shapes or legends could help overcome this issue. Participants highlighted the importance of integrating tactile displays into compact and discreet devices, such as phone-sized tools or wrist-mounted systems, to enhance accessibility and usability.

d) **Combination of Modalities:** Participants strongly advocated for hybrid systems that integrate audio, haptic, and tactile feedback to harness the unique advantages of each modality. They noted the Audio cues are excellent for delivering global awareness, such as navigation instructions or spatial descriptions, while haptic feedback provides precise, immediate alerts about path deviations or nearby obstacles. Tactile displays, on the other hand, are well-suited for tasks that require a detailed spatial understanding, such as route planning or pre-navigation exploration.

The adaptability of a combined system was seen as a major strength. For instance, audio feedback could be emphasized outdoors, where directional guidance and obstacle warnings are critical, whereas haptics feedback could be prioritized in noisy environments or situations that demand precise path corrections. Tactile displays could be used for pre-planning routes or exploring unfamiliar spaces. Participants stressed that the system should offer the flexibility to dynamically switch between modalities or allow users to customize their preferences based on the context.

### *Perceptions on Existing Assistive Technologies"*

a) **Common Challenges with Assistive Technologies:** Participants frequently mentioned the inaccuracies of GPS-based devices such as the Victor Reader Trek Deverell et al. (2020) and Trekker Breeze Roentgen et al. (2019), which were often off by 10–50 meters, limiting their reliability for precise navigation. These inaccuracies were particularly problematic when attempting to locate bus stops or building entrances. Overly complex user interfaces and excessive information were also cited as overwhelming, especially for those seeking straightforward navigation tools. Another significant challenge was the interference caused by continuous auditory feedback, which competed with participants' reliance on environmental sounds for spatial awareness. Devices that were physically cumbersome, such as those requiring users to hold them while managing other tasks, added to the burden. Additionally, existing devices failed to detect obstacles above waist height, such as low-hanging branches or signage, posing risks during navigation.

b) **Positive Experiences with Existing Technologies:** Despite the challenges, participants highlighted several technologies that offered valuable functionality. The Victor Reader Trek was particularly appreciated for its "walk mode" allowing users to virtually explore routes in advance, creating a mental map before setting out. Smartphone apps, including Seeing AI Granquist et al. (2021), BlindSquare Ponchillia et al. (2020), and VoiceVista Chang et al. (2024), were praised for their versatility in reading text, providing navigation guidance, and describing surroundings. The combination of Google Maps and BlindSquare offered useful walking instructions, proving useful in unfamiliar environments. Wearable devices like Envision glasses Gamage (2024) were noted for real-time OCR and environmental awareness, though their cost posed barriers to widespread adoption.

c) **Suggestions for Improvement:** Participants identified several improvements that could enhance the usability and adoption of assistive technologies. Customizable verbosity settings, allowing users to adjust the level and frequency of information provided, were highly recommended. Lightweight, wearable designs that integrate seamlessly into daily routines without drawing attention were emphasized. Enhanced obstacle detection, particularly for overhead and above waist hazards, was another common request.

Participants also expressed interest in tactile feedback as an alternative to auditory cues, especially in noisy environments. The integration of refreshable tactile displays for navigation could provide significant value. Extended battery life was noted as a critical feature for ensuring functionality during long journeys. Lastly, participants highlighted the need for simplified, intuitive interfaces to minimize cognitive load.

It was emphasized that individuals with visual impairments, like their sighted peers, strongly prefer the convenience of simply dressing up, grabbing their smartphone and cane, and heading out, without the need to spend time attaching additional gadgets to their bodies. Designs that require extra preparation or attachments are unlikely to gain wide acceptance among users with visual impairments. Training tools and tutorials were seen as essential for helping users adapt to new technologies.

d) **Emerging Needs and Desired Features:** The workshops also revealed broader needs, such as enhanced environmental awareness technologies capable of providing detailed spatial information, including the layout of rooms and the location of key features like entrances and elevators. Navigation in complex public spaces, such as parks, parking lots, or malls, emerged as a significant area where participants sought improvements.

## Discussion

### *Key Findings and Implications*

The relationship between visually impaired individuals and assistive technology reveals a persistent paradox: while traditional tools such as white canes and GPS-based navigation remain widely used, they fail to address critical gaps in usability and effectiveness. Workshop participants and survey respondents highlighted major limitations, particularly with GPS accuracy, which often lacks the precision needed for last-feet navigation. This shortcoming may make it difficult to locate bus stops, building entrances, and other crucial waypoints. Additionally, while auditory feedback is a valuable navigation aid, it frequently competes with important environmental sounds, leading to potential safety risks, particularly in urban settings where situational awareness is vital. The cognitive strain imposed by complex user interfaces further exacerbates usability challenges, with many assistive technologies proving too cumbersome for real-world navigation. Devices that require active handling pose additional burdens, as managing multiple objects, a smartphone, a white cane, and other personal belongings, limits mobility and discourages adoption.

Despite these obstacles, certain features of assistive devices have proven useful, such as the Victor Reader Trek's ”walk mode” for virtual route planning and smartphone applications that integrate navigation with real-time object recognition. However, a significant gap remains between the development of new assistive technologies and their widespread adoption. Many visually impaired individuals continue to rely on mainstream technologies, such as smartphones and general-purpose apps, rather than specialized assistive devices. The reasons for this disconnect are multifaceted: emerging tools often fail to offer compelling advantages over mainstream solutions, remain inaccessible due to cost, or require specialized training that is not readily available. The strong preference for lightweight, seamless, and wearable solutions underscores the need for assistive technologies that integrate naturally into daily life, minimizing setup and eliminating excessive user effort. If future innovations are to achieve widespread acceptance, they must prioritize real-time adaptability, ease of use, and multimodal feedback while addressing persistent gaps in spatial awareness, particularly in detecting above-cane obstacles and navigating complex environments. The failure of many new technologies to gain traction suggests that researchers must not only refine the technical performance of these devices but also rethink their accessibility, affordability, and training mechanisms to better align with the real-world needs of IVI.

Beyond the limitations of assistive technology, navigational challenges extend into broader issues of environmental design, public awareness, and urban planning. Above-waist obstacles, such as tree branches, protruding signage, and overhanging structures, remain undetectable by traditional white canes, posing significant risks to independent navigation. Additionally, inconsistencies in environmental cues, such as indistinct indoor layouts, irregular park pathways, and the lack of pedestrian-friendly routes in parking lots, further complicate navigation. Workshop participants emphasized that auditory complexities of urban noise, including construction, electric vehicles, and crowded public spaces, often disrupts navigation, highlighting the vulnerability of auditory-based strategies.

Confidence in navigation is closely tied to familiarity and mental mapping. Participants reported lower confidence levels when navigating unfamiliar environments, reinforcing the need for spatial awareness rather than simple path-following tools. The diversity of user needs further illustrates the shortcomings of one-size-fits-all solutions, while some users prioritize directional guidance, others require detailed environmental awareness, real-time updates on temporary obstacles, or adaptable levels of information delivery. Future assistive technologies must integrate multimodal feedback systems, allowing users to selectively engage different sensory channels based on context. Moreover, solutions must accommodate the dynamic nature of urban environments, where static navigation aids often fail in the presence of unexpected obstacles and evolving infrastructure.

A key consideration in the integration of assistive technologies is leveraging widely used platforms to facilitate adoption. Given the wide adoption of smartphones among IVI, participants emphasized that embedding assistive technologies into these devices could encourage widespread use. However, while smartphones offer a familiar and versatile interface, they introduce practical challenges that limit their effectiveness as primary navigation aids. Many users rely on them for route planning but find them impractical for real-time navigation due to the need to keep their hands free for safety, cane use, or guide dog handling. Concerns about dropping, misplacing, or having a smartphone stolen further complicate their suitability for on-the-go use. These insights suggest that while smartphones are valuable as an interface for assistive technology, hands-free or wearable solutions—such as smart glasses, wrist-worn devices, or bone-conduction audio systems, may provide a more practical approach to real-time navigation assistance. A similar logic was applied to traditional mobility aids, particularly the white cane. While some participants were open to integrating new technologies into familiar tools, concerns about weight, durability, and reliability created hesitancy. The white cane is an essential extension of spatial awareness, and modifications, such as embedded sensors or vibratory feedback, were seen as potentially disrupting rather than enhancing its primary function. Past attempts at smart canes were widely perceived as offering insufficient benefits to justify the added complexity. These responses highlight that for technological enhancements to be widely accepted, they must be lightweight, unobtrusive, and seamlessly integrated to preserve the fundamental usability of existing tools.

Finally, the financial constraints associated with assistive navigation technologies present a fundamental challenge to their widespread adoption. The survey responses highlight that a significant portion of visually impaired individuals spend little to nothing on dedicated assistive devices, relying instead on free smartphone apps and general-purpose tools. While some participants make periodic investments in low-to-mid-range solutions, such as cane replacements or screen readers, only a small fraction can afford specialized technologies. This reality underscores a critical barrier: the high cost of assistive technologies, compounded by the limited market size, makes them prohibitively expensive for many who could benefit from them the most. The perception that these tools are ”horrendously expensive” and often fail to deliver proportional value further discourages investment. However, the fact that over half of respondents were unwilling to specify a fixed budget for future technologies, instead stating that their willingness to pay would depend on perceived benefits, suggests that cost sensitivity is not absolute. If a truly effective and transformative assistive device were introduced, one that substantially enhances mobility, independence, and safety, users might be willing to stretch their financial limits to obtain it. This indicates that in addition to price considerations, developers should prioritize demonstrable impact and tangible advantages. Moreover, affordability strategies such as modular designs, flexible pricing models, or integration with widely available mainstream devices (like smartphones) could help bridge the financial gap. Ensuring that assistive technologies are not only functional but also accessible within realistic financial constraints is paramount to driving adoption and improving the daily lives of visually impaired individuals.

### *Limitations of the Methodology*

While our study provides insights into the navigation challenges and preferences of IVI, it is important to acknowledge certain limitations inherent in our methodology. First, the recruitment of participants primarily through the UC Davis Department of Ophthalmology and Vision Sciences and the Sacramento Society for the Blind and via word-of-mouth may introduce selection bias, as it potentially limits the diversity of the participant pool to those within specific networks or geographic locations. This could affect the generalizability of our findings to the wider visually impaired community. Additionally, the use of virtual workshops, although effective in facilitating discussion, and removing certain obstacles associated with travel, might constrain the depth of interaction compared to in-person settings. Other factors, such as internet connectivity and familiarity with virtual platforms, could influence the dynamics of the discussions and the quality of data collected. Another limitation of this study is the varying interpretations of key terms among participants; for example, some may consider smartphones or apps as assistive devices, while others view them as standard everyday technology. Finally, the self-reported nature of the survey responses, particularly concerning the level of visual impairment, could also lead to inaccuracies.

## Conclusions

This study highlights the persistent challenges visually impaired individuals face in navigating both indoor and outdoor spaces, emphasizing the need for assistive technologies that are accessible, intuitive, and seamlessly integrated into everyday life. While traditional tools such as white canes remain essential, our findings underscore the growing reliance on mainstream devices, particularly smartphones, over dedicated assistive tools. This preference suggests that future assistive technologies should prioritize integration with widely used platforms while addressing key usability concerns, including hands-free operation, multimodal feedback, and real-time adaptability.

The financial constraints associated with specialized assistive technologies further limit adoption, reinforcing the need for cost-effective solutions. Strategies such as leveraging existing hardware (e.g., smartphones), modular designs, and flexible pricing models could improve accessibility. Additionally, improving public infrastructure, such as better tactile markers and more consistent pedestrian pathways, could complement technological advancements, ensuring a more holistic approach to navigation support.

Ultimately, this study provides actionable insights for researchers, developers, and policymakers seeking to improve assistive navigation solutions. By prioritizing user-centered design, adaptability, and affordability, the next generation of assistive technologies can better empower visually impaired individuals to navigate with confidence and independence.

## Acknowledgements

The authors thank the participants for their valuable contributions and insights. We also thank the Sacramento Society for the Blind (https://societyfortheblind.org) for their assistance in recruiting participants. This work was supported in part by a seed grant from the Center for Information Technology Research in the Interest of Society and the Banatao Institute (CITRIS).

## References

Mohamed Bakali El Mohamadi, Adnan Anouzla, Nabila Zrira, and Khadija Ouazzani-Touhami. A systematic review on blind and visually impaired navigation systems. In Noredine Gherabi, Ali Ismail Awad, Anand Nayyar, and Mohamed Bahaj, editors, *Advances in Intelligent System and Smart Technologies*, pages 151–160. Springer International Publishing, 2024.

Mark S. Baldwin, Jennifer Mankoff, Bonnie Nardi, and Gillian Hayes. An activity centered approach to nonvisual computer interaction. *ACM Trans. Comput.-Hum. Interact.*, 27(2), March 2020. ISSN 1073-0516.

Raju Shrestha Bineeth Kuriakose and Frode Eika Sandnes. Tools and technologies for blind and visually impaired navigation support: A review. *IETE Technical Review*, 39(1):3–18, 2022.

Ruei-Che Chang, Yuxuan Liu, and Anhong Guo. Worldscribe: Towards context-aware live visual descriptions. In *Proceedings of the 37th Annual ACM Symposium on User Interface Software and Technology*, UIST '24, New York, NY, USA, 2024. Association for Computing Machinery. ISBN 9798400706288.

Lil Deverell, Joy Bhowmik, Bee Theng Lau, Abdullah Al Mahmud, Suku Sukunesan, Fakir M. A. Islam, and Denise Meyer. Use of technology by orientation and mobility professionals in australia and malaysia before covid-19. *Disability and Rehabilitation: Assistive Technology*, 17(3):260–267, 2020.

Mary Beatrice Dias, Ermine A. Teves, George J. Zimmerman, Hend K. Gedawy, Sarah M. Belousov, and M. Bernardine Dias. Indoor navigation challenges for visually impaired people. In Hassan A. Karimi, editor, *Indoor Wayfinding and Navigation*, pages 141–164. March 2015.

Fatma El-zahraa El-taher, Ayman Taha, Jane Courtney, and Susan Mckeever. A systematic review of urban navigation systems for visually impaired people. *Sensors*, 21(9), 2021.

Hugo Fernandes, Paulo Costa, Vitor Filipe, Hugo Paredes, and Joa˜o Barroso. A review of assistive spatial orientation and navigation technologies for the visually impaired. *Universal Access in the Information Society*, 18(1):155–168, 2019.

American Foundation for the Blind. Reviewing disability employment research for people who are blind or visually impaired, 2024. URL https://www.afb.org/research-and-initiatives/employment. Accessed: 2024-12-18.

Bhanuka Gamage. Ai-enabled smart glasses for people with severe vision impairments. *SIGACCESS Access. Comput.*, (137), March 2024. ISSN 1558-2337.

Christina Granquist, Sharon Y. Sun, Sandra R. Montezuma, Tony M. Tran, Robert Gage, and Gordon E. Legge. Evaluation

and comparison of artificial intelligence vision aids: Orcam myeye 1 and seeing ai. *Journal of Visual Impairment & Blindness*, 115(4):277–285, 2021.

Vahid Isazade. Advancement in navigation technologies and their potential for the visually impaired: a comprehensive review. *Spatial Information Research*, 31(5), 2023.

Gaylen Kapperman, Elisa Koster, and Rick Burman. The study of foreign languages by students who are blind using the jaws screen reader and a refreshable braille display. *Journal of Visual Impairment & Blindness*, 112(3):317–323, 2018.

Deborah Kendrick. Penfriend and touch memo: A comparison of labeling tools. *AccessWorld*, 12(9), 2011. URL https://www.afb.org/aw/12/9/15900. Accessed: 2025-04-04.

Gabriel Iluebe Okolo, Turke Althobaiti, and Naeem Ramzan. Assistive systems for visually impaired persons: Challenges and opportunities for navigation assistance. *Sensors*, 24(11), 2024.

OpenAI. ChatGPT (December 18 version), 2024. URL https://openai.com. Accessed: December 18, 2024.

LM Ortiz-Escobar, MA Chavarria, K Scho¨nenberger, S Hurst, MA Stein, A Mugeere, and M Rivas Velarde. Assessing the implementation of user-centred design standards on assistive technology for persons with visual impairments: a systematic review. *Frontiers in Rehabilitation Sciences*, 4, September 6 2023.

Paul E. Ponchillia, Sang-Jin Jo, Kendra Casey, and Stephanie Harding. Developing an indoor navigation application: Identifying the needs and preferences of users who are visually impaired. *Journal of Visual Impairment & Blindness*, 114(5): 344–355, 2020.

Alec Radford, Jong Wook Kim, Tao Xu, Greg Brockman, Christine McLeavey, and Ilya Sutskever. Robust speech recognition via large-scale weak supervision. In *Proceedings of the 40th International Conference on Machine Learning*, ICML'23. JMLR.org, 2023.

Fabiana Sofia Ricci, Lorenzo Liguori, Eduardo Palermo, John-Ross Rizzo, and Maurizio Porfiri. Navigation training for persons with visual disability through multisensory assistive technology: Mixed methods experimental study. *JMIR Rehabilitation and Assistive Technologies*, 11, 2024.

U. R. Roentgen, G. J. Gelderblom, and L. P. de Witte. Users' evaluations of four electronic travel aids aimed at navigation for persons who are visually impaired. *Journal of Visual Impairment & Blindness*, 105(10):612–623, 2019. doi: 10.1177/0145482X1110501008. Originally published 2011, republished in 2019.

Ami Rokach, David Berman, and Alison Rose. Loneliness of the blind and the visually impaired. *Frontiers in Psychology*, 12, 2021.

M.F. Saaid, A.M. Mohammad, and M.S.A. Megat Ali. Smart cane with range notification for blind people. In *2016 IEEE International Conference on Automatic Control and Intelligent Systems (I2CACIS)*, pages 225–229, 2016. doi: 10.1109/I2CACIS.2016.7885319.

Michael Waisbourd, Omar M. Ahmed, Jonathan Newman, Manishi Sahu, Daniel Robinson, Lamia Siam, Carisa B. Reamer, Tong Zhan, Michelle Goldstein, Sara Kurtz, Marlene R. Moster, Lisa A. Hark, and L. Jay Katz. The effect of an innovative vision simulator (orcam) on quality of life in patients with glaucoma. *Journal of Visual Impairment & Blindness*, 113(4): 332–340, 2019.

WeWALK. Wewalk smart cane – smart cane for the visually impaired. https://wewalk.io/en/, 2025. Accessed: 2025-04-05.

C. Wickramaarachchi, R. Jayathilaka, T. Suraweera, S. Thelijjagoda, L. Kollure, T. Liyanage, et al. Can visual impairment impact your income potential? *PLoS ONE*, 18(4):e0284553, 2023. doi: 10.1371/journal.pone.0284553.